\documentstyle[psfig,12pt]{article}
\newcommand{\AOP}[3]{Annals\ Phys.\ {\bf {#1}}, {#2} ({#3})}
\newcommand{\CMP}[3]{Comm.\ Math.\ Phys.\ {\bf {#1}}, {#2} ({#3})}

\newcommand{\PRD}[3]{Phys.\ Rev.\ D {\bf {#1}}, {#2} ({#3})}

\newcommand{\NPB}[3]{Nucl.\ Phys.\ B {\bf {#1}}, {#2} ({#3})}
\newcommand{\NPS}[3]{Nucl.\ Phys. \ Proc. \ Suppl. {\bf {#1}}, {#2} ({#3})}
\newcommand{\PLB}[3]{Phys.\ Lett.\ B {\bf {#1}}, {#2} ({#3})}
\newcommand{\RMP}[3]{Rev.\ Mod.\ Phys.\  {\bf {#1}}, {#2} ({#3})}

\newcommand{\SNP}[3]{Sov.\ J.\ Nucl.\ Phys.\ {\bf {#1}}, {#2} ({#3})}

\newcommand{\beq}{\begin{equation}}
\newcommand{\eeq}{\end{equation}}
\newcommand{\be}{\begin{eqnarray}}
\newcommand{\ee}{\end{eqnarray}}
\newcommand{\ba}{\begin{array}}
\newcommand{\ea}{\end{array}}
\newcommand{\ur}[1]{(\ref{#1})}

\newcommand{\eq}[1]{Eq.~(\ref{#1})}
\newcommand{\eqs}[2]{Eqs.(\ref{#1}, \ref{#2})}
\newcommand{\Eq}[1]{Eq.~(\ref{#1})}

  \newcommand{\la}[1]{\label{#1}}

  \def\beqr{\begin{eqnarray}}
  \def\eeqr{\end{eqnarray}}

\begin{document}
\thispagestyle{empty}
%\begin{flushright}
\hspace{11.4cm }LU TP 99-27\\
\hspace*{12.0cm }NORDITA-1999/57 HE \\
\hspace*{12.0cm}hep-th/9909078\\
\hspace*{12.0cm }\today
%\end{flushright}
\vspace{0.2cm}

\vskip 3true  cm

\begin{center}
{\Large\bf On Statistical Mechanics of Instantons}\\ \vskip0.5cm
{\Large\bf in the $CP^{N_c-1}$ Model}
\vskip 2true cm

{\large\bf
Dmitri Diakonov$^{\diamond *}$ and Martin Maul$^{\diamond \dagger}$}
\\
\vskip 1true cm
\noindent
{\it
$^\diamond $NORDITA, Blegdamsvej 17, 2100 Copenhagen \O, Denmark \\
\vskip .2true cm
$^*$Petersburg Nuclear Physics Institute, Gatchina,
St.~Petersburg 188 350, Russia
\vskip .2true cm
$^\dagger$ Theoretical Physics II, Lund University, S-223 62 Lund, Sweden}
\end{center}

\vskip 2.5true cm
\begin{abstract}
\noindent
We introduce an explicit form of the multi-instanton weight
including also
instanton--anti-instanton interactions for arbitrary $N_c$ in the
two-dimensional $CP^{N_c-1}$ model. To that end, we use
the parametrization of multi-instantons in terms of instanton
`constituents' which we call `zindons' for short. We study
the statistical mechanics of zindons analytically (by means
of the Debye-H\"uckel approximation) and numerically (running
a Metropolis algorithm). Though the zindon parametrization
allows for a complete `melting' of instantons we find that,
through a combination of dynamical and purely
geometric factors, a dominant portion of topological
charge is residing in well-separated instantons and
anti-instantons.
\end{abstract}
\newpage
\section{Introduction}
In the last two decades there have been much evidence that
the specific fluctuations of the
gluon field carrying topological charge, called instantons, play a very
important role in the dynamics of QCD, the theory of strong
interactions \cite{Ne98,SS98}. Instantons are most probably
responsible for one of the main
features of strong interactions, namely for the spontaneous breaking
of chiral symmetry \cite{DP86}.
Whether they play a role in the confinement property of QCD is, however,
not clear.
The difficulty in addressing this problem may be connected to the fact
that, strictly speaking, the  multi-instanton solution in the vacuum
is not known.
Instead, in a dilute-gas approximation, one treats the instanton
vacuum as a superposition of one-instanton solutions. For large
distances the one-instanton gluon field falls
off as $A_\mu(x) \sim 1/x^3$, which means that correlations between
instantons at large distances are vanishing in the dilute-gas picture.
However, the confinement phenomenon by itself seems to
indicate that certain correlations of gluon fields do not fall off
when the distance is increased. Therefore, if instantons are of any
relevance to confinement, it cannot be seen in the dilute-gas approximation:
one has to enrich the arsenal of instanton methods.
\newline
\newline
This motivates us to study a system  where
the multi-instanton solution is known, where one can find many
features of the strong interaction, and which is simple enough
to allow for analytic solutions in some limits. Such a system
is provided by the two-dimensional $CP^{N_c-1}$ model.
This model is known to possess both asymptotic freedom and instantons.
Contrary to the $d=4$ Yang--Mills theory where the
{\em multi}-instanton solutions are not available in an explicit form,
in the $CP^{N_c-1}$ model they are known explicitly, and for any
`number of colors' $N_c$ \cite{GP,ALV78}. Moreover, the multi-instanton
weight arising from integrating out quantum oscillations about
instantons is also known \cite{FFS79a,FFS79b,BL79}. In
addition, the model is exactly solvable at large $N_c$
\cite{ALV78,Wi79}, and the spectrum is known for all $N_c$
\cite{FR86}. Therefore, the model is well
suited to monitor instantons in a controllable way. It should be added
that the model has been much studied by lattice simulations (see,
e.g. \cite{Vi96,Vi92,Vi92a}), and instantons have been identified there by a
`cooling' procedure \cite{MS94}.
\newline
\newline
Since the pioneering paper of Fateev, Frolov and Schwartz
\cite{FFS79a} it has been known that the multi-instanton system
in the $O(3)$ sigma model resembles that of a two-dimensional Coulomb
plasma; it can be exactly bosonized to produce a sine-Gordon theory. In
a true vacuum, however, one has both instantons and anti-instantons;
such a system can be bosonized and solved exactly only at a specific
value of the coupling \cite{BuLi} which may not
be realistic.  Therefore, the general case still remains a
difficult and unsolved dynamical problem.
\newline
\newline
In Sec.~\ref{sec2} we overview the $CP^{N_c-1}$ model.
In Sec.~\ref{sec3} we derive the general partition function for the
$CP^{N_c-1}$ in terms of a special form of instanton
variables, which are called `zindons'.
In Sec.~\ref{sec4} we
shall describe an approximation to the partition function called the
Debye-H\"uckel approximation, which is capable to account
for many features of the
instanton ensemble.  However, that approximation is in general
insufficient.  Therefore, in this paper we combine analytical study of
the instanton--anti-instanton ensemble with a numerical simulation.  In
Sec.~\ref{sec5} we describe the Metropolis algorithm used to study the
behavior of the model in Monte Carlo simulations.  It will turn out
that, strictly speaking, the instanton ensemble
 does not exist for $N_c=2$ and that a
regularization procedure is required to yield meaningful results.  In
Sec.~\ref{sec6} we consider the question whether the system has a
thermodynamically stable
density, i.e., whether the average number of instantons and anti-instantons
grows proportional to the volume of the system. Finally in Sec.~\ref{sec7}
we have a look at the size distribution, both seen from the instanton
input and from  the topological charge density.
\section{The $CP^{N_c-1}$ model}
\label{sec2}
For a number of colors $N_c$ the main objects of the $CP^{N_c-1}$
model are $N_c$ complex fields $v_A(x), A=1,\dots, N_c$. Defining the
norm $|v|^2 = \sum_{A=1}^{N_c} |v_A|^2$ one can introduce
the normalized fields $u_A$, a projector P and the unitary operator G
via:
\[
u_A = v_A/|v|, \qquad P_{AB} = u_A u^*_B, \qquad G = 1-2P,
\]
\begin{equation}
P^2=P=P^\dagger,\qquad G=G^\dagger,\qquad G^2=1.
\end{equation}
One can introduce a vector potential $A_\mu$ and a covariant derivative
$\nabla_\mu$:
\begin{equation}
A_\mu = \frac{i}{2}( u_A \partial_\mu u^*_A-u^*_A \partial_\mu u_A),
\qquad \nabla_\mu = \partial_\mu - i A_\mu, \qquad (\mu = 1,2)\;.
\end{equation}
$A_\mu$ is a real field, depending on the x-space variables $x$ and $y$
that can be identified with the complex variables $z=x+iy$ and $z^*=x-iy$.
The field strength tensor and the topological charge density are:
\begin{equation}
F_{\mu\nu} = \partial_\mu A_\nu - \partial_\nu A_\mu,\;\;\;\;\;\;
q(x)=\frac{1}{4\pi}\epsilon_{\mu\nu}F_{\mu\nu}
=\frac{-i}{2\pi}\epsilon_{\mu\nu}\partial_\mu u_A^*\partial_\nu u_A.
\end{equation}
The theory is determined by the action
\beq
S = \int d^2x \sum_{A=1}^{N_c}| \nabla_\mu u_A |^2 = \frac{1}{8}
\int d^2 x {\rm Tr} \left[ \partial_\mu G \partial_\nu G\right],
\la{action}\eeq
and the partition function is:
\[
{\cal Z}=\int\!Du_A(x)Du^*_A(x)DA_\mu(x)\delta(|u|^2-1)
\exp\left(-\frac{1}{g^2}
\int\!d^2x|\nabla_\mu u_A|^2\right)
\]
\beq
=\int\!Du_A(x)Du^*_A(x)\delta(|u|^2-1)
\exp\left(-\frac{1}{8g^2}\int\!d^2x
{\rm Tr}[\partial_\mu G \partial_\nu G]\right).
\la{partfun}\eeq
This formulates a nonlinear theory of self-interacting fields, where
the nonlinearity is forced by the condition $|u| = 1$.
There is a topological charge $Q$ in the theory defined as:
\beq
Q = \int\! d^2 x\,q(x) = \frac{1}{2\pi}\oint dx_\mu A_\mu
= \frac{1}{4\pi} \int\! d^2 x\, \epsilon_{\mu\nu} F_{\mu\nu} =
\frac{i}{16\pi} \int\! d^2x\, \epsilon_{\mu\nu}\, {\rm Tr} \left[
G\partial_\mu G\partial_\nu G\right].
\la{topcharge}\eeq
Because of its unitarity $G$ has the property $G\partial G = -\partial G G$,
which can be used to show:

\beq
0 \leq \int d^2x {\rm Tr}\left[
\left(\partial_\mu G \pm i\epsilon_{\mu\nu} G\partial_\nu G \right)
\left(\partial_\mu G \pm i\epsilon_{\mu\nu} G\partial_\nu G \right)
\right] = 16 (S\mp 2\pi Q) \;.
\eeq
This tells that the minimal action for a given topological charge of
a field configuration, i.e. a solution of the equation of motion, is
obtained if the field satisfies the self-duality equation:
\beq
\partial_\mu G \pm i\epsilon_{\mu\nu} G\partial_\nu G
= 0\;\; \Leftrightarrow\;\; \partial_\mu v \mp i\epsilon_{\mu\nu}
\partial_\nu v =0.
\la{selfdual}\eeq
Introducing the complex derivative $\partial_z = (\partial_x
-i\partial_y)/2$, it means that one finds two types of solutions:
\beq
\partial_{z^*} v= 0\; \longleftrightarrow \;S=2\pi Q
\la{sd}\eeq
and
\beq
\partial_{z} v = 0\; \longleftrightarrow\;S=-2\pi Q \;,
\la{asd}\eeq
which are the Cauchy--Riemann conditions \cite{BP75}.  One calls the
first an instanton solution and the second one an anti-instanton
solution.
\section{`Zindons'}
\label{sec3}
A general multi-instanton solution of the self-duality \eq{sd}
with topological charge $Q=N_+$ is a product of monomials in $z$
\cite{FFS79a}:
\beq
v^{inst}_A = c_A \prod_{i=1}^{N_+}(z-a_{Ai}),\;\;\;\;\;
A=1,...,N_c.
\eeq
Similarly, a general multi anti-instanton solution of \eq{asd} with
topological charge $Q=-N_-$ is a product of monomials in the complex
conjugate variable $z^*$:
\beq
v_A^{anti} = c^\prime_A \prod_{j=1}^{N_-}(z^*-b^*_{Aj}),\;\;\;\;\;
A=1,...,N_c.
\eeq
For a general configuration with $N_+$ instantons and $N_-$ anti-instantons
we shall consider a product Ansatz \cite{BuLi}:
\beq
v_A =  \prod_{i=1}^{N_+} (z-a_{Ai}) \prod_{j=1}^{N_-} (z^*-b^*_{Aj}),
\la{Ansatz}\eeq
where $a_{iA}$ and $b_{jA}$ are fixed 2-dim points written as complex
numbers. Those points have been called `instanton quarks' or
`instanton constituents' in the past. We suggest a shorter term
{\em zindon} to denote these entities, in analogy with similar objects
introduced in the Yang--Mills theory \cite{DP}. `Zindon' is Tajik or
Persian word for `castle' or `prison'; they are of relevance to
describe confinement in the model. There are, thus, $N_c$ types or `colors'
of the instanton zindons (denoted by $a_A$) and $N_c$ types of
anti-instanton zindons or anti-zindons (denoted by $b_A$).
\newline
\newline
The action and the topological charge for such
a multi-instanton--anti-instanton Ansatz are
\[
\left.\begin{array}{c}S\\2\pi Q\end{array}\right\}
=2\int d^2x \left[
\frac{\sum_A|v_A|^2|\sigma_A|^2
-\left|\sum_A |v_A|^2\sigma_A\right|^2/|v|^2}{|v|^2}
\right.
\]
\[
\left.\pm \:\frac{\sum_A|v_A|^2|\tau_A|^2
-\left|\sum_A |v_A|^2\tau_A\right|^2/|v|^2}{|v|^2}
\right],
\]
\beq
\sigma_A=\sum_{i=1}^{N_+}\frac{1}{z-a_{Ai}},\qquad
\tau_A=\sum_{j=1}^{N_-}\frac{1}{z^*-b^*_{Aj}},\qquad
|v|^2=\sum_{A=1}^{N_c}|v_A|^2.
\label{acttop}\eeq
The topological charge can be immediately found from the Cauchy theorem:
\beq
Q = \frac{1}{2\pi}\oint dx_\mu A_\mu = N_+ -N_-\,.
\eeq
The geometric meaning of the zindon coordinates becomes especially clear
if one takes a single-instanton solution, $v_A(z)=c_A(z-a_A)$. Putting
it into the action Eq.~\ur{acttop} one gets the standard form of the
instanton profile,
\beq
S =2\pi Q =2\int\! d^2x \frac{\rho^2}{[(x-x_0)^2 + \rho^2]^2} =2\pi,
\la{singleinst}\eeq
where the instanton center $x_0$ is given by the {\em center
of masses} of zindons,
\beq
x_{0\mu}=\frac{\sum_{A=1}^{N_c}|c_A|^2a_{A\mu}}
{\sum_{A=1}^{N_c}|c_A|^2},
\la{center}\eeq
while the spread of the field called the instanton size $\rho$ is
given by the spatial {\em dispersion} of zindons comprising the instanton,
\beq
\rho^2=\frac{\sum_{A=1}^{N_c}|c_A|^2|a_A-x_0|^2}
{\sum_{A=1}^{N_c}|c_A|^2}
=\frac{\sum_{A<B}|c_A|^2|c_B|^2|a_A-a_B|^2}
{\left(\sum_A|c_A|^2\right)^2}.
\la{size}\eeq
In \eqs{center}{size} all coordinates $x_0,\,a_A$ are understood as
2-dimensional vectors.
In the multi-instanton case the action and topological charge densities
will have extrema not necessarily coinciding with the position of the
zindon center of masses. Only when a group of zindons of all $N_c$
colors happens to be well spatially isolated from all the rest zindons,
in the vicinity of this group one can speak about getting a classical
instanton profile \ur{singleinst} with the collective coordinates given
by \eqs{center}{size}, the role of the coefficients $c_A$ played by
a product of separations between a given zindon $a_{Ai}$ and all the rest
zindons and anti-zindons,
\beq
|c_A|^2=\prod_{i^\prime\neq i}^{N_+}|a_{Ai}-a_{Ai^\prime}|^2
\prod_{j=1}^{N_-}|a_{Ai}-b_{Aj}|^2.
\la{cA}\eeq
As a matter of fact, it means that there is no need in introducing
{\em explicitly} extra degrees of freedom $c_A$. It should be added that
in the thermodynamic limit,
with the system volume $V$ going to infinity, we expect the number
of zindons to be proportional to  $V$. Meanwhile, the number of
the coefficients $c_A$ is fixed (equal to $N_c$). Therefore,
one can neglect this degree of freedom as it plays no role
in the thermodynamic limit; we shall therefore put all $c_A=1$.
\newline\newline
The quantum determinants in the multi-instanton background, defining
the instanton weight or the measure of integration over collective
coordinates $a_{Ai}$, has been computed in Ref. \cite{FFS79a} in case
of $N_c=2$ (which is the $O(3)$ sigma model) and in Refs.
\cite{FFS79b,BL79} for arbitrary $CP^{N_c-1}$; the results of these
references coincide up to notations.
\subsection{Multi-instanton weight in the case $N_c=2$}
In the $N_c=2$ case  the multi-instanton weight
is that of the Coulomb gas \cite{FFS79a,BL79}:
\beq
{\cal D}a\cdot w_a = \prod_{A=1}^{N_c=2}\prod_{i=1}^{N_+}d^2a_{Ai}\cdot
\frac{\prod_{A=1,2}\prod_{i<i^\prime}|a_{Ai}-a_{Ai^\prime}|^2}
{\prod_{i,i^\prime}|a_{1i}-a_{2i^\prime}|^2}
\left[\left(\frac{2\pi}{g^2}\Lambda\right)^2 C(2)\right]^{N_+}
\la{wa}\eeq
where $\Lambda$ is the renormalization-invariant combination of the
bare charge $g^2$ and of the UV cutoff $M$ needed to regularize the
theory; it is the analog of $\Lambda_{QCD}$. The numerical coefficient
$C(2)$ depends on the regularization scheme used; in the Pauli--Villars
scheme it has been found in Ref. \cite{Jev} to be
\beq
C(2)=\left(\frac{2}{e}\right)^6\frac{3}{\pi^2}=0.04822082\,.
\la{C2}\eeq
We shall absorb this constant in the definition of $\Lambda$.
In the one-loop approximation the charge $g^2$ in \eq{wa} remains
unrenormalized; in two loops it starts to run. It is interesting that
for accidental reasons (the number of zero modes per unit
topological charge, equal four, coinciding with the ratio of the
second-to-first Gell-Mann--Low coefficients, equal to
$b^\prime/b=4N_c/N_c=4$) the coupling $g^2$ does not `run' in the
two-loop approximation either:  $2\pi/g^2$ in \eq{wa} should be just
replaced by $(b/2)^{b^\prime/2b}=1$.
For anti-instanton zindons $b_{Aj}$ one has a similar expression,
with $N_+$ replaced by $N_-$. We denote the corresponding weight by
${\cal D}b\cdot w_b$.
\subsection{Partition function for arbitrary $N_c$}
\label{sec3.2}

\begin{figure}
\centerline{\psfig{figure=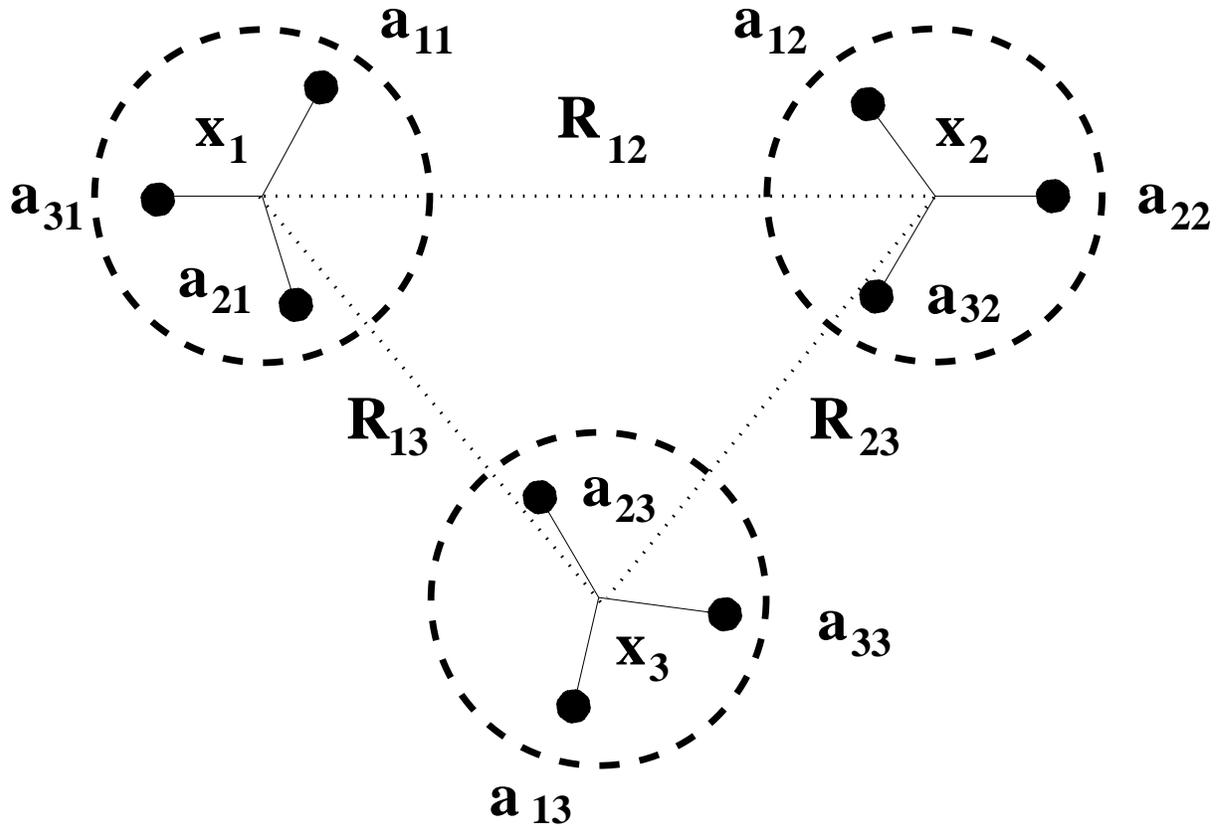,width=16cm}}
\caption{The `dilute' regime for $N_c=3$:
Concentrating around the center-of-mass points $x_1$, $x_2$,
and $x_3$ zindons of three different colors are closely grouped together
to form  instantons with large distances $R_{ij}$ between them.
The configuration corresponds to a dilute gas of non-overlapping
instantons.}
\label{cluster}
\end{figure}

The general expression for
the multi-instanton weight $w_a$ in the $CP^{N_c-1}$ model is
\cite{FFS79b,BL79}:
\begin{eqnarray}
w_a &=& \exp\Bigg[ \sum_{i<j}^{N_+}\sum_A^{N_c} \ln |a_{Ai}-a_{Aj}|^2
- \frac{N_c}{8\pi} \int d^2 x \ln |v_A|^2 \partial ^2 \ln |v_A|^2
\nonumber \\
&& -\frac{N_c}{2} N_+ \ln \left(\sum_{A=1}^{N_c} |c_A|^2 \right)
+\frac{N_c}{2}N_+ + N_+  \sum_{A=1}^{N_c} \ln |c_A|^2 \Bigg]\;,
\nonumber \\
v_A &=& \prod_{i=1}^{N_+}c_A(z-a_{Ai}) \;.
\la{waint}
\end{eqnarray}
In the following we set $|c_A|=1$ for reasons explained above.
The second (integral) term in \eq{waint} can be analytically
computed only for $N_c=2$; it gives the Coulomb interaction of
\eq{wa}. For $N_c > 2$
we approximate the integral term by a simpler explicit expression.
To that end, let us consider a
configuration with zindons of $N_c$ colors
forming clusters well separated from
each other, see Fig.~\ref{cluster}. We denote the centers of clusters
by $x_i$,
\begin{equation}
x_i = \frac{1}{N_c} \sum_{A=1}^{N_c} a_{Ai}\;,
\end{equation}
the sizes of the instantons associated with those clusters by $\rho_i$,
\begin{equation}
\rho_i^2 = \frac{1}{N_c}\sum_{A=1}^{N_c}(a_{Ai}-x_i)^2
         = \frac{1}{N_c^2}\sum_{A<B} (a_{Ai}-a_{Bi})^2\;,
\end{equation}
and the separation between clusters by
\begin{equation}
R_{ij} = |x_i - x_j|\;,
\end{equation}
assuming $R_{ij}^2\gg \rho_{i,j}^2$.
In such geometry the integral term can be
easily evaluated, yielding
\begin{eqnarray}
I\{a_{Ai}\} &=&
-\frac{N_c}{8\pi} \int d^2x  \ln |v_A|^2 \partial ^2 \ln |v_A|^2
\nonumber \\
&\simeq&
-\frac{N_c}{2} \Bigg[
2\sum_{i<j}^{N_+} \ln R_{ij}^2+
\sum_{i=1}^{N_+} \ln \rho_i^2
+ N_+(\ln N_c+1)
\Bigg] + {\cal O}\left(\frac{\rho^2_{i,\;j}}{{R_{ij}}^2}\right)\;.
\label{above}
\end{eqnarray}
We would like to write the functional of zindon
positions $I\{a_{Ai}\}$ in such a way that it
\begin{itemize}
\item[(a)] depends only on the separations
between individual zindons, $a_{Ai}-a_{Bj}\;$;
\item[(b)] is symmetric
under permutations of same-color zindons, $(a_{Ai}\leftrightarrow a_{Aj})$;
\item[(c)] reduces to \eq{above} in the dilute regime;
\item[(d)] at $N_c=2$ comes to the exact expression valid for any geometry:
\end{itemize}
\begin{equation}
I^{N_c=2}\{a_{Ai}\} =
-\left[ \sum_{i,j=1}^{N_+} \ln(a_{1i}-a_{2j})^2 + N_+(1-\ln 2)\right]\;.
\end{equation}
The solution to this problem is, naturally, not unique.
We shall suggest two forms of approximate
expressions for $I\{a_{Ai}\}$, both satisfying the above requirements a-d.
The first form is
\begin{eqnarray}
I\{a_{Ai}\} &=& - \frac{N_c}{2} \sum_{i,j=1}^{N_+}
             \ln\left[\sum_{A<B}(a_{Ai}-a_{Bj})^2\right]
           + \frac{N_c}{2}N_+(\ln N_c -1)
\nonumber \\
&& + N_c\frac{N_+(N_+-1)}{2} \ln \frac{N_c(N_c-1)}{2}\;.
\end{eqnarray}
Combining it with other terms in (\ref{waint}) we get the following
multi-instanton weight:
\begin{eqnarray}
w_a &=& \exp \beta \Bigg[\sum_{i<i'}^{N_+} \sum_{A=1}^{N_c}
\ln(a_{Ai}-a_{Ai'})^2 \Lambda^2
-\frac{N_c}{2}\sum_{i,i'=1}^{N_+} \ln
\left[ \sum_{A<B}^{N_c}(a_{Ai}-a_{Bi'})^2 \Lambda^2 \right]
\nonumber \\
&& + N_c \frac{N_+(N_+-1)}{2}\ln \frac{N_c(N_c-1)}{2} \Bigg]\;.
\label{form1}
\end{eqnarray}
This form of the interactions between zindons belonging to
instantons will be used in computer simulations of the ensemble,
see below. This form is, however, inconvenient for analytical
estimates since it involves many-body interactions
represented by $\ln \sum_{A<B}(a_{Ai}-a_{Bi'})^2$. Therefore, we suggest a more simple second form involving
only two-body interactions:
\begin{equation}
\tilde I\{a_{Ai}\} = -\frac{1}{N_c-1} \sum_{i,j=1}^{N_+} \sum_{A<B}^{N_c}
\ln(a_{Ai}-a_{Bj})^2 + \frac{N_c}{2}N_+(\ln N_c -1)\;.
\end{equation}
The second form reproduces correctly only the leading term in
(\ref{waint})
in the `dilute' regime; however it is exact for $N_c=2$. Consequently,
the multi-instanton weight can be written in a compact form:
\begin{equation}
\tilde w_a = \exp \left\{ \frac{\beta}{2}
\frac{N_c}{N_c-1}
\sum^{N_+}_{i,j=1}
\sum^{N_c}_{A,B=1}
{\cal P}_{AB} \ln(a_{Ai}-a_{Bj})^2\Lambda^2
\right\}
\label{form2}
\end{equation}
Here ${\cal P}_{AB}$ is a projector $N_c\times N_c$ matrix such that
${\cal P}_{AB} = (N_c-1)/N_c$ when $A=B$ and ${\cal P}_{AB}=-1/N_c$ when $A\neq B$, so
that ${\cal P}^2={\cal P}$, and ${\cal P}$ has $N_c-1$ unit eigenvalues and one zero.
\newline
\newline
In both forms, (\ref{form1}) and (\ref{form2}), we have introduced an
`inverse temperature' $\beta$ for future convenience, in fact $\beta=1$.
We have also borrowed certain powers of the scale $\Lambda$ from
the overall weight coefficient to make the arguments
of the logarithms dimensionless. A comparison of numerical
simulations using the weight (\ref{form1}) with analytical calculations
using the weight (\ref{form2}) shows that the results do not
differ significantly; it means that ensembles generated with the two
weights are to a good accuracy equivalent. At $N_c=2$
they both coincide with the exact weight (\ref{wa}). For the
multi--anti-instanton weight $w_b$ one can use either of the forms
(\ref{form1}), (\ref{form2}) with an obvious substitution
of the instanton-zindon coordinates $a_{Ai}$ by anti-zindon
coordinates $b_{Ai}$.
\newline
\newline
Finally, we turn to the instanton--anti-instanton
interaction factor $w_{ab}$.
We assume that it is mainly formed
by the classical defect of the action,
\beq
w_{ab}=\exp\left(-\frac{1}{g^2}U_{int}(a,b)\right),\;\;\;\;\;
U_{int}(a,b)=S_{Ansatz}-2\pi(N_++N_-),
\la{wab1}\eeq
where $S_{Ansatz}$ is the action computed on the product Ansatz,
\ur{Ansatz}. Again, it is possible to evaluate this quantity in a
regime, when zindons belonging to definite instantons or anti-instantons
are grouped together and are well separated from other groups. After
some work one obtains the classical interaction potential between
zindons and anti-zindons in a form of `dipole interactions' in two
dimensions, valid for any number of colors $N_c$:
\[
U_{int}=8\pi\sum_{all\;I\bar I\;pairs}\;\;
\frac{1}{N_c}\sum_A(a_A-x_a)_\mu (b_A-x_b)_\nu
\frac{1}{R^2}\left(\delta_{\mu\nu}-2\frac{R_\mu R_\nu}{R^2}\right),
\]
\beq
x_a=\frac{a_1+...+a_{N_c}}{N_c},\;\;\;\;\;
x_b=\frac{b_1+...+b_{N_c}}{N_c},\;\;\;\;\;
R_\mu=x_{a\mu}-x_{b\mu}.
\la{Uint1}\eeq
In terms of zindons, this is a many-body interaction, however,
\eq{Uint1} can be rewritten as a pair interaction of individual
zindons, which, naturally, appears to be of a Coulomb type. This is one of the
great advantages of using the zindon parametrization of instantons.
We rewrite the dipole interaction \ur{Uint1} as
a sum of two-body interactions using the $N_c\times N_c$ projector
${\cal P}_{AB}$ introduced above:
\beq
U_{int}=-\frac{4\pi}{N_c}\sum_{i=1}^{N_+}\sum_{j=1}^{N_-}\sum_{A,B=1}^{N_c}
{\cal P}_{AB}\ln|a_{Ai}-b_{Bj}|^2.
\la{Uint2}\eeq
In the case when groups of $N_c$ zindons form clusters well separated
from groups of $N_c$ anti-zindons, this reduces to the dipole interaction
(\ref{Uint1}).
Notice that the interaction of ``same-color'' zindons is repulsive
while that of ``different-color'' is attractive.
At $N_c=2$  the projector matrix
${\cal P}_{AB}$ is
\beq
{\cal P}_{AB}^{N_c=2}=\frac{1}{2}\left(
\begin{array}{cc}1&-1\\-1&1\end{array}\right),
\la{PAB2}\eeq
so that the interaction becomes just a sum of Coulomb interactions
between charges of different signs. In this case this form has been
known previously \cite{Jev,BuLi}. In Ref. \cite{BuLi} it has been shown
that this form of the interaction possesses an additional conformal
symmetry, and one can think that its domain of validity in the zindon
configuration space is wider than the one where it has been
actually derived. We shall, thus, use for the mixed
instanton--anti-instanton weight:
\begin{equation}
w_{ab} = \exp \left\{ 2\beta \beta_1
\sum_{i=1}^{N_+}  \sum_{j=1}^{N_-} \sum_{A,B=1}^{N_c}
{\cal P}_{AB} \ln \left[ (a_{Ai}-b_{Bj})^2 \Lambda^2 \right]\right\}, \quad
\beta_1 = \frac{2\pi}{g^2 N_c}\;.
\label{wab}
\end{equation}
The full partition function takes the form
(for arbitrary instanton angle $\theta$):
\begin{equation}
Z = \sum_{N_+,N_-}
\frac{e^{i\theta N_+}}{(N_+!)^{N_c}}
\frac{e^{i\theta N_-}}{(N_-!)^{N_c}}
\int {\cal D}a {\cal D}b \Lambda^{2N_c(N_++N_-)} w_a w_b w_{ab}
\label{partfu1}
\end{equation}
with $w_{ab}$ given by (\ref{wab}), and $w_a$,$w_b$ given either by
(\ref{form1}) or (\ref{form2}). If one uses the form
(\ref{form2}) the interpretation is especially clear:
the partition function describes two systems of zindons
(instanton and anti-instanton ones) experiencing logarithmic Coulomb
interactions, whose strength is $N_c-1$ times stronger
for same-color zindons than for  different-color zindons.
One has attraction for zindons--anti-zindons of different
color and repulsion for zindons--anti-zindons of the same color.
At $N_c=2$ one can think of the ensemble as of that of
$e^+,e^-,\mu^+,\mu^-$ particles
\cite{BuLi}. The interaction of opposite-kind
zindons are suppressed by an additional factor $\beta_1 = 2\pi/(g^2N_c)$.
It is actually a running coupling, depending on the scale in the problem.
Since there is only one scale here, the density of zindons at the
thermodynamic equilibrium, this $\beta_1$ should be found self-consistently
from the arising equilibrium density at the end of the calculations.
\section{The Debye-H\"uckel approximation}
\label{sec4}
To get an insight in the dynamics of the system by simple analytical
methods we consider here the so-called Debye-H\"uckel approximation to the
partition function \ur{partfu1}.
Let us consider first a general case of a
statistical ensemble of $F$ kinds of particles with two-body interactions
given by:
\beq
U_{\rm int} = \sum_{i\le j} u_{\rm int}(x_i-x_j)\;.
\eeq
The number of particles of kind $f$ is $N_f, f=1,\dots,F$. One
writes the partition function for fixed numbers of particles as:
\beq
Z = e^{-\beta {\cal F}} = \frac{1}{N_1!\dots N_F!} \int
\prod_{f=1}^F \prod_{i_f=1}^{N_f} dx ^{f}_{i_f} e^{-\beta U_{\rm int}}
\;,
\eeq
where ${\cal F}$ is the Helmholtz free energy and $x^{f}$ are coordinates
of particles of kind $f$. The one-particle density of particles of kind $f$
is defined as:
\be n_f(x_1^{f}) &=& \frac{1}{Z} \frac{1}{N_1!\cdots N_F!}
\int d x_2^f \cdots dx_{N_f}^f \prod_{g \neq f}
\prod_{i_g=1}^{N_g} dx_{i_g}^g e^{-\beta U_{\rm int}}\;,
\nonumber \\
\int dx_1^f n_f(x_1^f) &=& 1\;,
\ee
which differs from the full partition function by that one
does not integrate over the coordinates of one particle of
kind $f$. Similarly, one can introduce the two-particle density
by avoiding integrating over coordinates of particle $f$ and
of particle $g$:
\beq
n_{fg}(x_1^f,x_2^g) = \frac{1}{Z} \frac{1}{N_1!\cdots N_F!}
\int dx_2^f\dots d x_{N_f}^f \: dx_1^g dx_3^g \cdots dx_{N_g}^g
\prod_{h\neq f \atop h \neq g} \prod_{i_h=1}^{N_h} dx_{i_h}^h
e^{-\beta U_{\rm int}} \;.
\eeq
This function can be written as
\beq
n_{fg}(x_1^f,x_2^g) = n_f(x_1^f)n_g(x_2^g)
e^{-\beta\omega_{fg}(x_1^f-x_2^g)}
\;,
\eeq
which serves as a definition of the two-particle correlation function
$\omega_{fg}(x_1,x_2)$.
If this function is zero the particles are not correlated.
Introducing, in the similar fashion, higher-order densities, one
can derive an (infinite)
chain of equation for those densities. The so-called correlation energy is:
\beq
E_{\rm corr} =- \frac{1}{2} \sum_{fg} \frac{N_fN_g}{V^2}
\int \int dx_1 dx_2 u_{fg}(x_1-x_2) e^{-\beta \omega_{fg}(x_1-x_2)} \;.
\la{gen1}
\eeq
Knowing the correlation energy as function of temperature one can restore the
Helmholtz free energy from the general relation:
\beq
E_{\rm corr} = \frac{\partial}{\partial \beta}(\beta {\cal F}) \;.
\la{gen2}
\eeq
The Debye-H\"uckel approximation consists in assuming that $\beta \omega_{fg}$
is small and linearizing the equations \cite{LLbook}. It is usually
justified, when the temperature is high, $\beta \rightarrow 0$. In the
Debye-H\"uckel approximation it is also convenient to pass to
the Fourier transforms of the pair interaction potential $u_{fg}$ and of the
correlation function $\omega_{fg}$. The basic equations from where
one finds the
correlation functions take the form:
\beq
\omega_{fg}(k) = u_{fg}(k) + \beta\sum_h
\frac{N_h}{2V} [
u_{gh}(-k) \omega_{fh}(k) + u_{fh}(k) \omega_{gh}(-k)] \;.
\eeq
In our case the particles are zindons belonging either
to instantons or to anti-instantons.
Let us denote $u_{A_1A_2},\omega_{A_1 A_2}$ the
interaction and the correlation of zindons of color
$A_1$, $A_2$ belonging to instantons. Similarly,
$u_{\bar B_1 \bar B_2},\omega_{\bar B_1 \bar B_2}$ refer to zindons
belonging to anti-instantons. and
$u_{A \bar B},\omega_{ A \bar B}$ refer to instanton zindons of
color $A$ and anti-instanton zindons of color $\bar B$.
The approximation is applicable only if all interactions are
two-body. For that reason we choose to work with the instanton weights given
by Eqs.~(\ref{form2}), (\ref{wab}):
\begin{eqnarray}
{\cal Z}&=&\sum_{N_+,\,N_-}\frac{e^{i\theta N_+}}{(N_+!)^{N_c}}
\frac{e^{-i\theta N_-}}{(N_-!)^{N_c}}\int {\cal D}a{\cal D}b\;
\Lambda^{2 N_c (N_++N_-)}\; \tilde w_a \tilde w_b w_{ab}\;.
%\nonumber \\
%\tilde w_a &=& \exp \beta\Bigg[\sum_{A=1}^{N_c}\sum_{i<i^\prime}
%\ln(|a_{Ai}-a_{Ai^\prime}|^2 \Lambda^2)-
%\frac{N_c}{2}
%\sum_{i,\;i^\prime}
%\sum_{A>B}\ln \left(|a_{Ai} - a_{Bi^\prime}|^2\Lambda^2 \right)
%\nonumber \\
%\tilde w_b &=& \exp \beta\Bigg[\sum_{B=j}^{N_c}\sum_{j<j^\prime}
%\ln(|b_{Bj}-b_{Bj^\prime}|^2 \Lambda^2)-
%\frac{N_c}{2}
%\sum_{j,\;j^\prime}
%\sum_{A>B}\ln\left( | b_{Aj} - b_{Bj^\prime}|^2\Lambda^2 \right)
%\nonumber \\
%\la{partfu2}
\end{eqnarray}
Since our interactions are all of the logarithmic type the
Fourier transforms of $u_{fg}$ can be written as:
\begin{eqnarray}
u_{A_1 A_2}(k) &=& v_{A_1 A_2}\frac{4\pi}{k^2}, \quad \;
v_{A_1 A_1} = 1, \quad v_{A_1 A_2} = -\frac{1}{N_c-1}
 \quad {\rm if}\; A_1 \neq A_2,
\nonumber \\
u_{\bar B_1 \bar B_2}(k) &=& v_{\bar B_1 \bar B_2}\frac{4\pi}{k^2}, \quad
v_{\bar B_1 \bar B_1} = 1, \quad v_{\bar B_1 \bar B_2} = -\frac{1}{N_c-1}
\quad {\rm if}\; \bar B_1 \neq \bar B_2,
\nonumber \\
u_{A \bar B}(k) &=& v_{A \bar B}\frac{4\pi}{k^2}, \quad
v_{A \bar A} = 2\beta_1\frac{N_c-1}{N_c}, \quad
v_{A \bar B} = - 2 \beta_1 \frac{1}{N_c}
 \quad {\rm if}\; A \neq B \;.
\end{eqnarray}
It is now convenient to change the notation for the $\omega$
correlation functions to:
\begin{equation}
\omega_{AA} = \omega_1,\;\;
\omega_{AA' \atop A\neq A'} = \omega_2,\;\;
\omega_{\bar B \bar B} = \omega_3,\;\;
\omega_{\bar B \bar B' \atop \bar B \neq \bar B'}
= \omega_4,\;\;
\omega_{A\bar B} = \omega_5,\;\;
\omega_{A\bar B' \atop \bar B' \neq A} = \omega_6.
\end{equation}
The above relations for the $v$'s induce similar relations
for the $\omega$'s:
\begin{equation}
\omega_2 = -\frac{1}{N_c-1}\omega_1,\;
\omega_4 = -\frac{1}{N_c-1}\omega_3,\;
\omega_6 = -\frac{1}{N_c-1}\omega_5\;.
\end{equation}
With the new notations we get the following set of linear
equations for the three independent correlation functions $\omega_{1,2,3}$:
\begin{eqnarray}
\beta \omega_1 &=& -\gamma[1+\tilde n_+ \beta\omega_1
                     +\beta' \tilde n_- \beta\omega_5]\;,
\nonumber \\
\beta\omega_3  &=&  -\gamma[1+\tilde n_- \beta\omega_3
                     +\beta' \tilde n_+ \beta\omega_5]\;,
\nonumber \\
\beta \omega_5 &=& -\gamma[\beta' +
\frac{1}{2} \beta'\beta(\tilde n_+\omega_1 + \tilde n_- \omega_3)
+ \frac{1}{2}\beta\omega_5 (\tilde n_+ +\tilde  n_-)],
\end{eqnarray}
where
\begin{eqnarray}
\gamma &=& \frac{4\pi\beta}{k^2},\quad
\beta' = 2\beta_1 \frac{N_c-1}{N_c}= \frac{4\pi(N_c-1)}{g^2N_c^2} \;.
\nonumber \\
\tilde n_+    &=& \frac{N_c}{N_c-1}\frac{N_+}{V}
              = \frac{N_c}{N_c-1}n_+ \;, \quad
\tilde n_-    = \frac{N_c}{N_c-1}\frac{N_-}{V}
              = \frac{N_c}{N_c-1}n_-\;.
\end{eqnarray}
We get the following solutions for the set of linear equations:
\begin{eqnarray}
\omega_1 &=& \frac{-4 \pi [k^2 + 4 \pi \beta \tilde n_-(1-\beta'^2)]}
                  {(k^2+\kappa_1^2)(k^2+\kappa_2^2)} \;,
\nonumber \\
\omega_3 &=& \frac{-4 \pi [k^2 + 4 \pi \beta \tilde n_+(1-\beta'^2)]}
                  {(k^2+\kappa_1^2)(k^2+\kappa_2^2)}\;,
\nonumber \\
\omega_5 &=& \frac{-4 \pi  \beta' k^2}
                    {(k^2+\kappa_1^2)(k^2+\kappa_2^2)} \;,
\nonumber \\
\kappa_{1,2} &=& 2\pi\beta \left[(\tilde n_++\tilde n_-)
\pm \sqrt{(\tilde n_+-\tilde n_-)^2 + 4 \tilde n_+\tilde n_-\beta'^2}\;\right] \;.
\la{deb}\end{eqnarray}
Returning to the x-space we observe that all correlation functions
are exponentially decreasing at large separations between zindons,
as contrasted to the original logarithmically growing interactions.
This phenomenon is usually referred to as `Debye screening'; in our
case it takes the particular form of \eq{deb}.
\newline
\newline
Using Eqs.~(\ref{gen1},\ref{gen2}) one can reconstruct the partition function
for a fixed number of instantons and anti-instantons:
\begin{eqnarray}
Z(N_+,N_-) &=& \exp
\left\{- \frac{\beta N_c^2}{4(N_c-1)}V
\left[
\frac{\tilde n_+ +\tilde n_-}{2}\left(
\ln \frac{\tilde \Lambda^2}
{4\pi\beta \sqrt{\tilde n_+\tilde n_-(1-\beta'^2)}}
+1\right)\right.\right.
\nonumber \\ &&
\left.\left.
-\frac{1}{4}\sqrt{(\tilde n_+-\tilde n_-)^2
+ 4\tilde n_+ \tilde n_-\beta'^2}\;
\ln \frac{\kappa_1}{\kappa_2} \right]\right\}
\frac{(V \Lambda^2)^{N_c(N_++N_-)}}{(N_+!)^{N_c}(N_-!)^{N_c}},
\nonumber \\
\tilde \Lambda &=& 2\Lambda  e^{-\gamma_E} \;.
\end{eqnarray}
At $N_+$=$N_-$ = $N/2$ corresponding to the instanton angle $\theta =0$
it simplifies to:
\begin{eqnarray}
Z(N) &=& \exp \Bigg[-
\frac{\beta N N_c^3}{8(N_c-1)^2}
\Bigg\{ 1 - \ln \left(
\frac{2 \pi \beta N_c N}{\tilde \Lambda^2 (N_c-1)V}
\right)
\nonumber \\
&&   -\frac{1}{2} \ln[(1-\beta')(1+\beta')]- \frac{1}{2}\beta'
\ln \frac{1-\beta'}{1+\beta'}
\Bigg\} \Bigg]
\frac{(V \Lambda^2)^{N_c N}}{\left(\frac{N}{2} !\right)^{2N_c}}
\;.
\end{eqnarray}
Finally, one can find the optimal density of instantons and anti-instantons
by maximizing the above partition function in $N$.
For a fixed $\beta'$ and $\Lambda$ we get then the following density
of instantons plus anti-instantons:
\begin{equation}
n = (n_- + n_+) = \frac{<N>}{V \Lambda^2}
= \Bigg[
(1+\beta')^\frac{1+\beta'}{2}
(1-\beta')^\frac{1-\beta'}{2}
4 \pi \beta
\frac{e^{2\gamma_E}}{8}
\frac{N_c}{N_c-1}2^{\frac{8(N_c-1)^2}{\beta N_c^2}} \Bigg]
^\frac{\beta N_c^2}{8(N_c-1)^2 -\beta N_c^2}\;.
\label{densdeb}
\end{equation}
Notice that the density is stable in the thermodynamic limit
$V \rightarrow \infty$.
At $|\beta'|=1$
corresponding to the value of the instanton--anti-instanton interaction
constant $\beta_1=\pm\frac{N_c}{2(N_c-1)}$ the system becomes unstable.
This could be anticipated from direct inspection of (\ref{partfu1}):
at that value of $\beta_1$ the attraction between zindons of different
color belonging to instantons and to anti-instantons exceeds the
repulsion between same-kind zindons, and the system collapses.
\newline
\newline
It should be stressed that all quantities involved (like the instanton
density $n$, the screening masses $\kappa_{1,2}$ etc.) are stable in the
limit $N_c\to\infty$. This should be contrasted to the old conjecture
by Witten \cite{Wi79} that instantons die out in the large $N_c$ limit.
Though this conjecture has been criticized later on \cite{Jev2,Polbook}
it remains a widely-believed prejudice.
\section{The Metropolis algorithm and regularization procedure}
\label{sec5}
\begin{figure}
\centerline{\psfig{figure=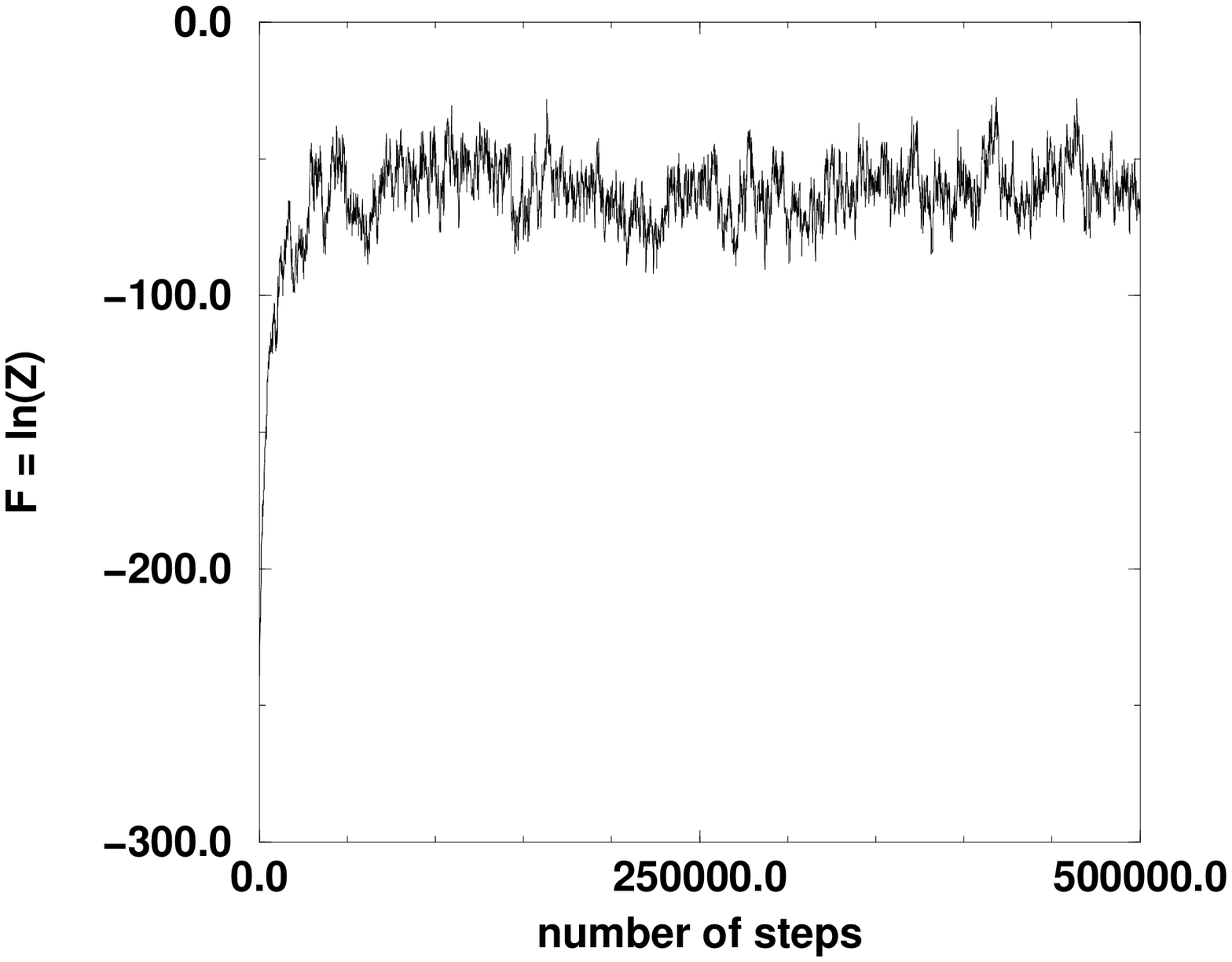,width=8cm}
            \psfig{figure=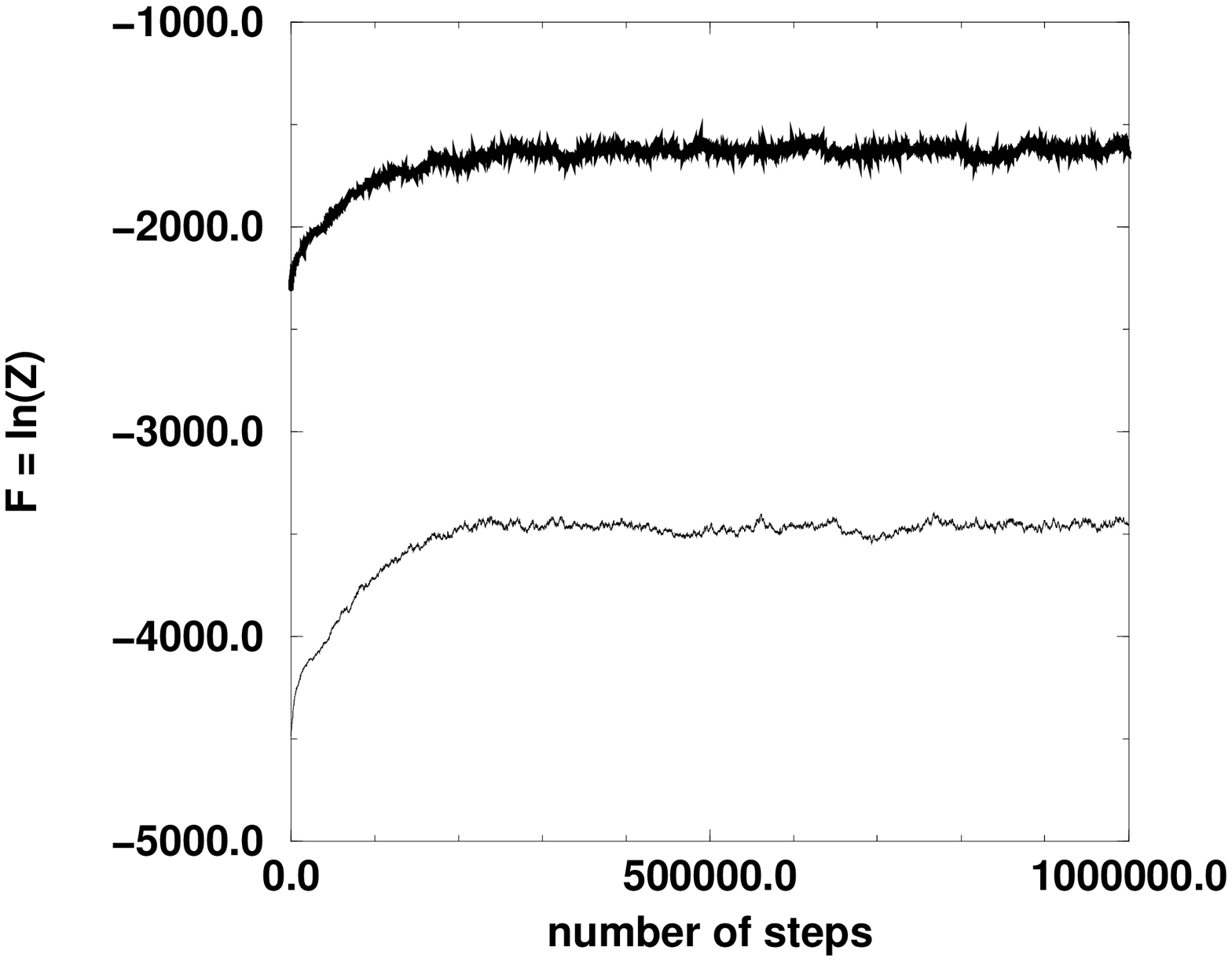,width=8cm}}
\caption{Stability of the Metropolis algorithm for $N_c=2$
[${\epsilon'}^2=0.001$] (on the
left) and $N_c=3$ (on the right, bold  line)
          $N_c=4$ (on the right, thin  line). The graphs
show the free energy improved step by step by the Metropolis algorithm until
a plateau of `important' configurations is reached. In all three runs
we used  $N_+=30$ and $L=10$.
In all cases the interaction between instantons and anti-instantons
is switched off $(\beta_1 = 0)$.}
\label{metropolis}
\end{figure}

\begin{figure}

\centerline{\psfig{figure=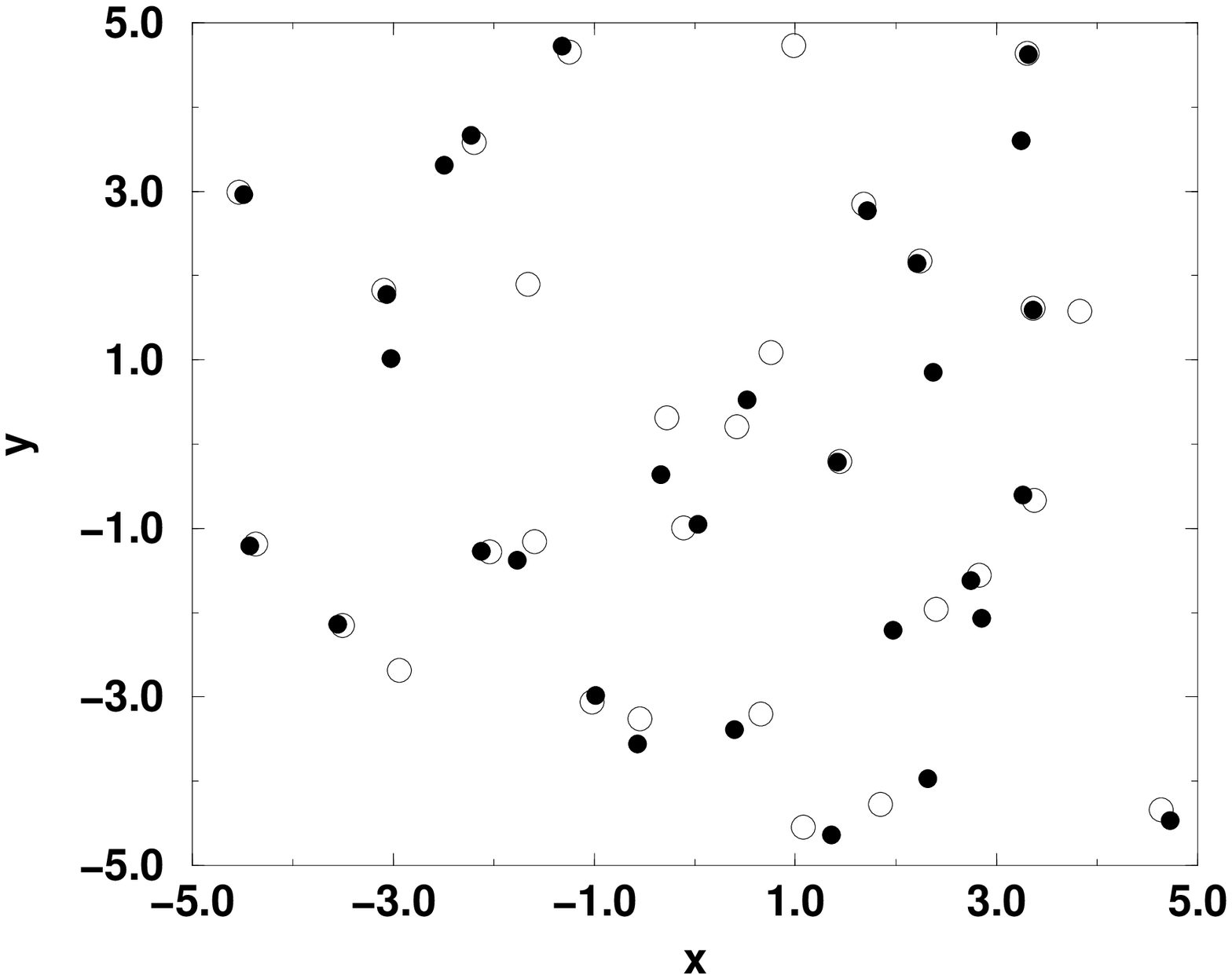,width=8cm}}
\centerline{\psfig{figure=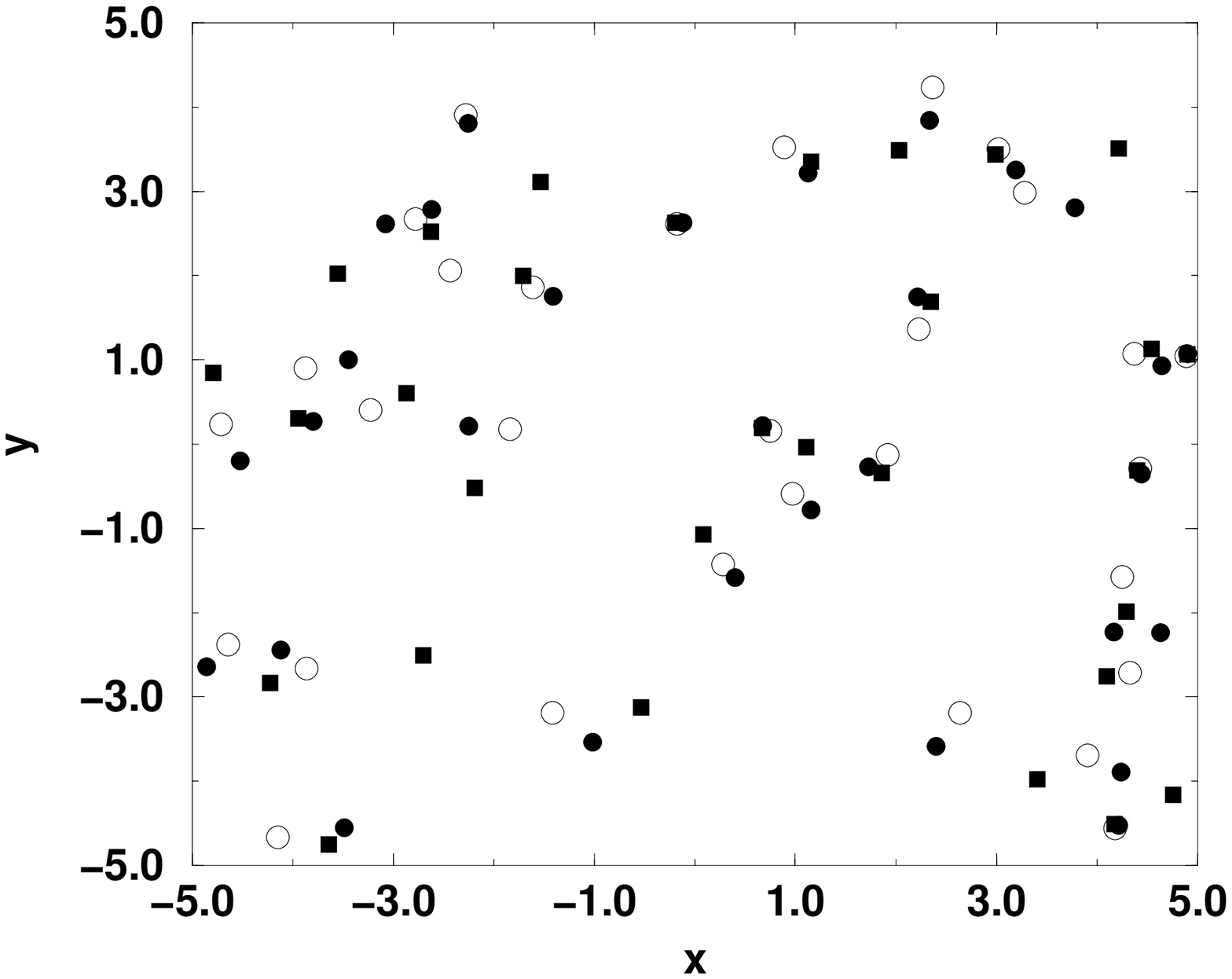,width=8cm}
            \psfig{figure=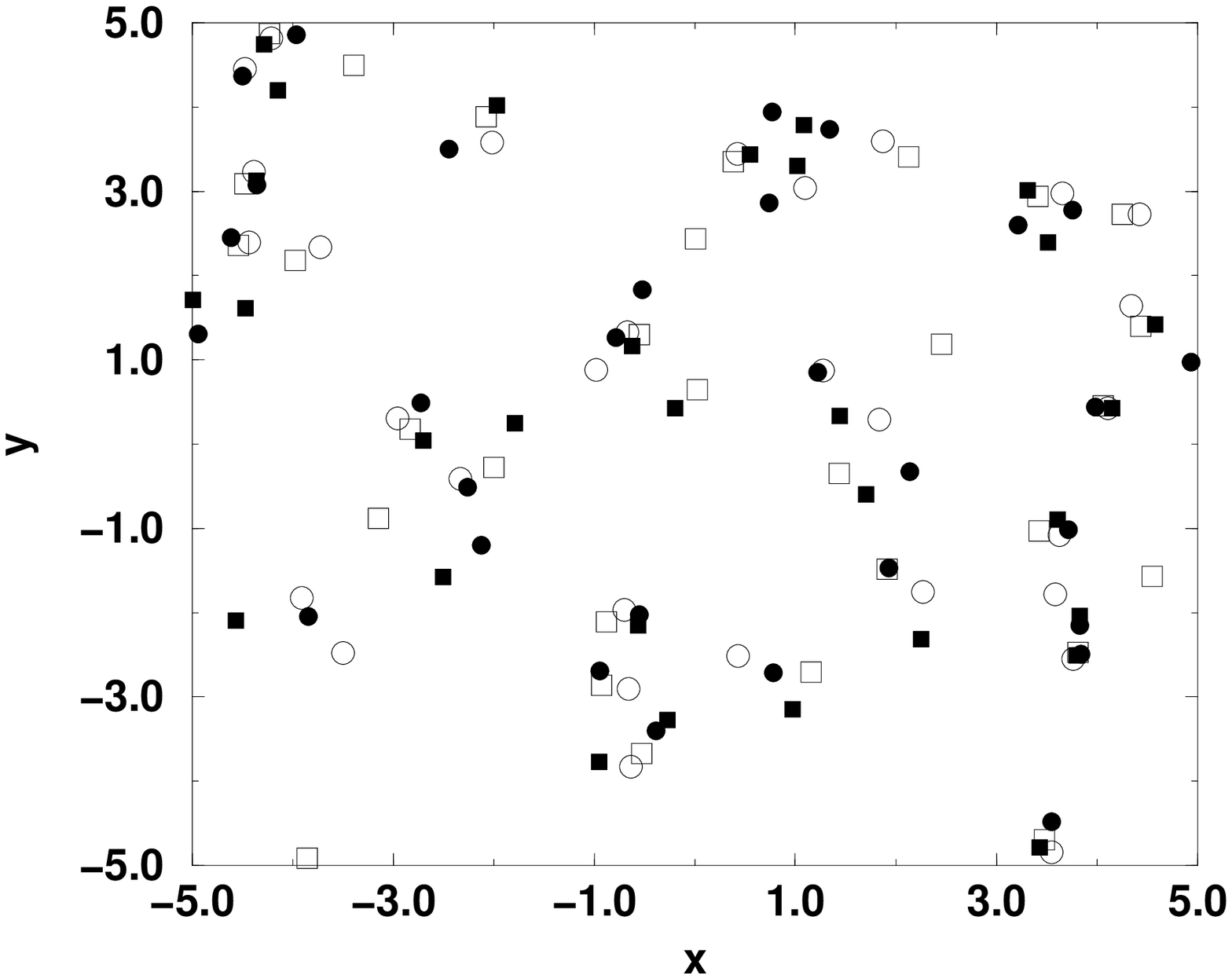,width=8cm}}
\caption{Snapshots of zindon configurations at the plateau after
500.000 Metropolis steps: The figure shows the spatial
distribution of zindons
for $N_c=2$ (on the top ), $N_c=3$ (bottom left) and $N_c=4$
(bottom right). The filled and empty circles and squares  symbolize
different colors. Only for $N_c=2$ a condensation into
neutral pairs is observed;
for higher $N_c$ the free energy is finite though many distinct
`color-neutral' $N_c$-plets are seen. Only instanton-zindons are shown.}
\label{snapshot}
\end{figure}

\begin{figure}
\centerline{\psfig{figure=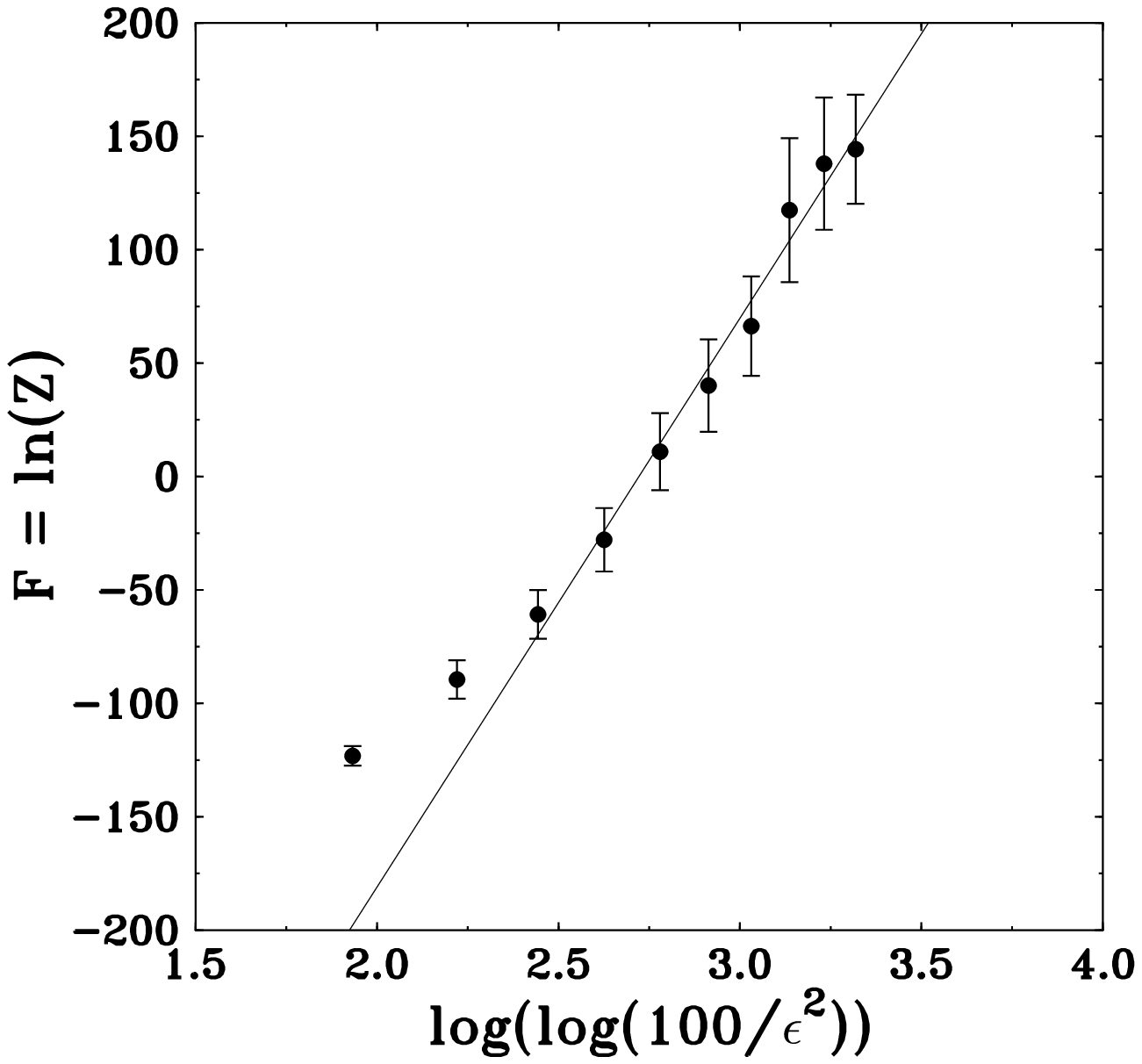,width=8cm}}
\centerline{\psfig{figure=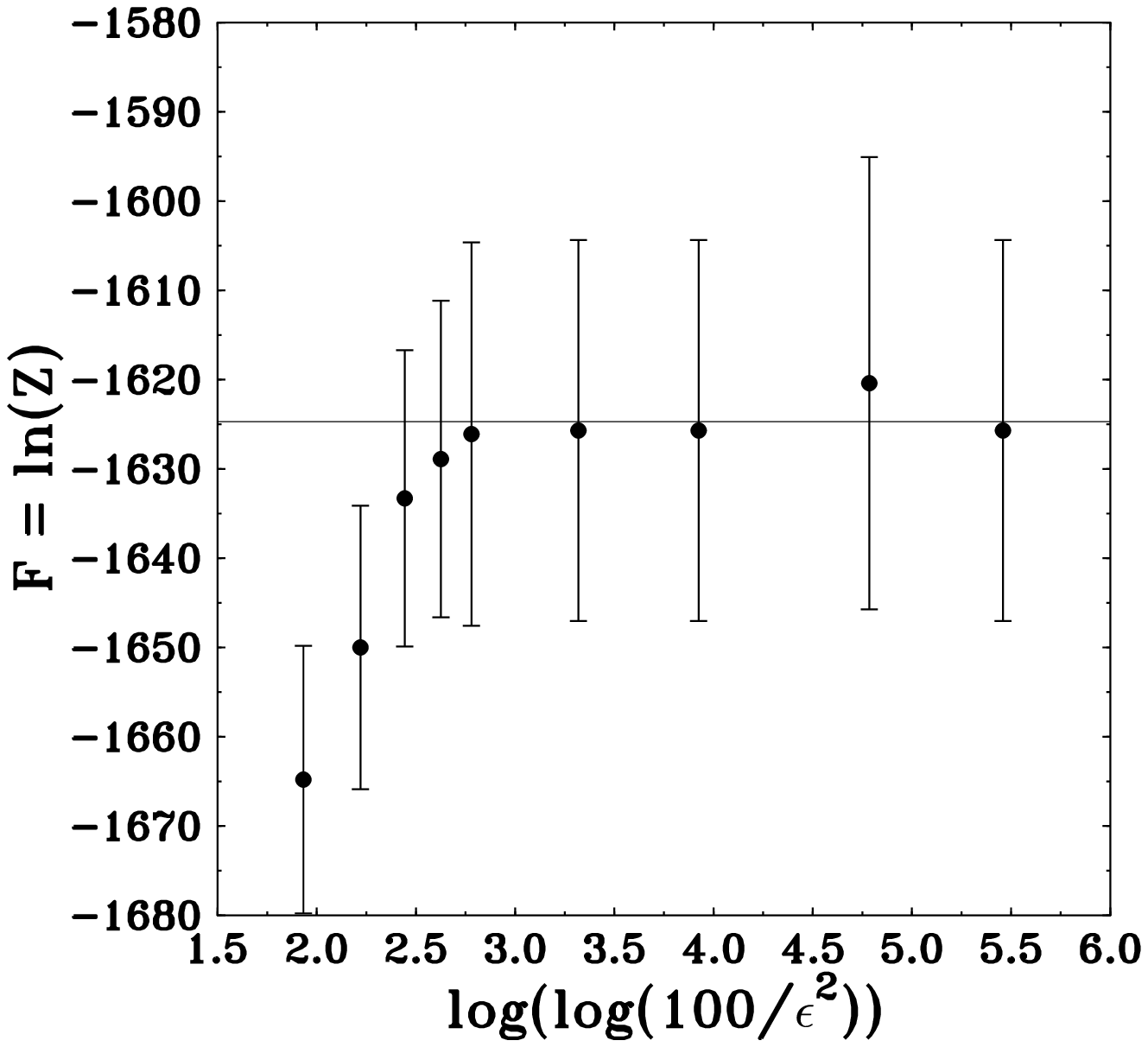,width=8cm}
            \psfig{figure=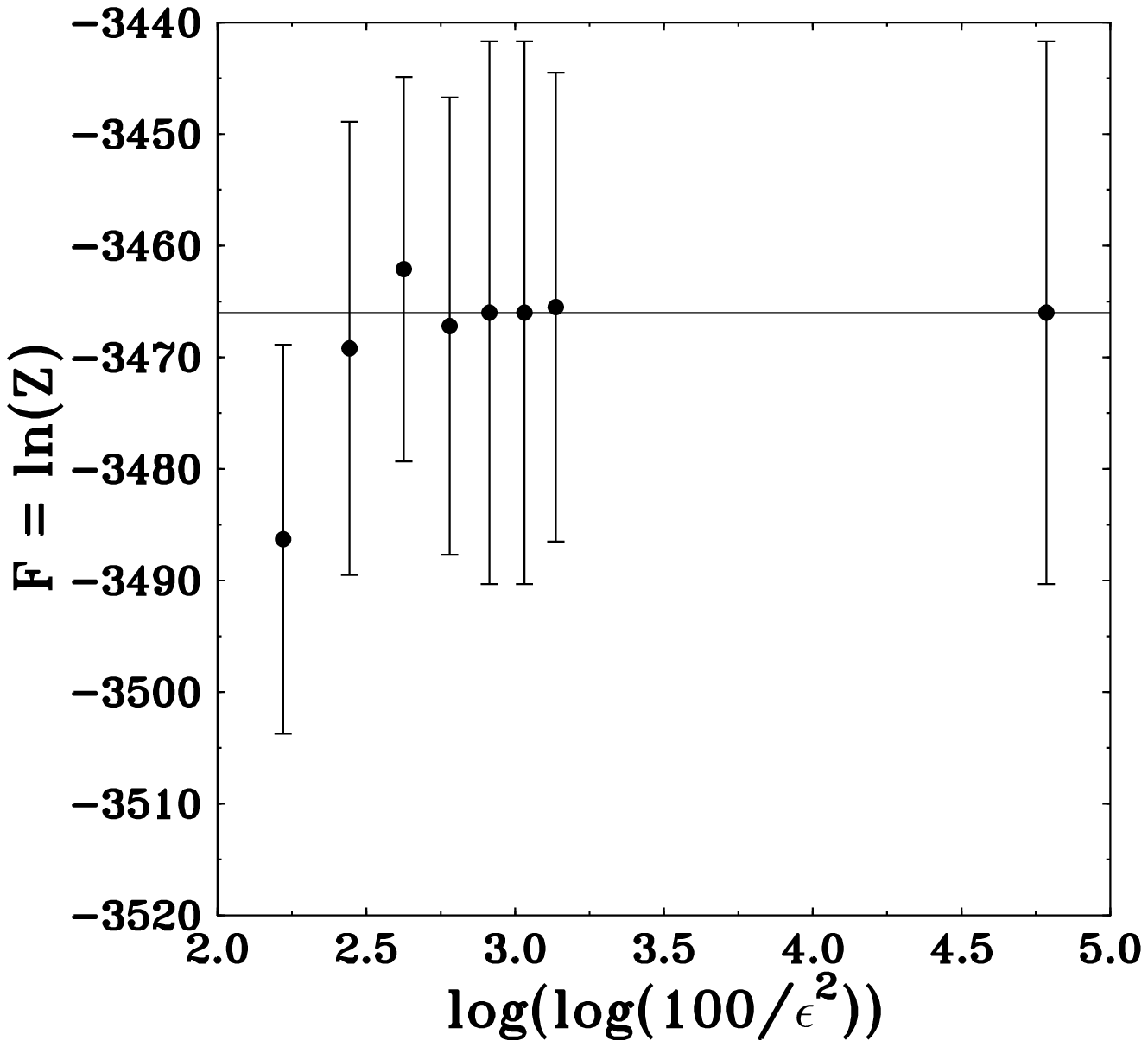,width=8cm}}
\caption{Divergent/convergent behavior of the free energy:
The free energy
is plotted versus $\ln(\ln(100/\epsilon))$ for $N_c=2$
(top), $N_c=3$ (bottom left), and $N_c=4$ (bottom right).
One observes a divergent behavior for $N_c=2$, while the free energy is
finite for $N_c=3$ and $N_c=4$. The last point in the plots with
 $N_c=3$ and $N_c=4$ is calculated using $\epsilon=0$.}
\label{divergence}
\end{figure}

The numerical treatment of instantons in the $CP^{N_c-1}$ model is problematic
for $N_c=2$, where the theory is equal to the two-dimensional $O(3)$
model. The partition function in this case involves a product
of integrals of the type $\int d \lambda^2 /\lambda^2$ and is therefore
ultraviolet divergent.  There are two ways of imposing a regularization.
The first consists in decreasing the powers in the interactions
(\ref{wa}). Physically,
this is equivalent to the introduction of a
temperature $T\equiv 1/\beta$, see Eqs.~(\ref{form1}),(\ref{form2}), and
(\ref{wab}).
We reach the physical limit by taking  $\beta \rightarrow 1$. For small
values of $\beta$ we expect the model to be consistent with the predictions
of the Debye-H\"uckel approximation.
The second possibility is using an ultraviolet cutoff $\epsilon$.
Technically, this can be easiest done by changing
the logarithmic interaction between zindons:
\begin{equation}
\ln(a_{Ai} - a_{Ai'})^2
\rightarrow \ln\left[(a_{Ai} - a_{Ai'})^2 +\epsilon^2\right],
\quad {\rm etc}\;.
\end{equation}
In both cases the physical limit $T\rightarrow 1$ and $\epsilon \rightarrow 0$
does not exist for the free energy, however one can get meaningful
results for physical quantities that are connected to derivatives
of the free energy. An unpleasant point is that this $\epsilon$ regularization
destroys the scaling properties of the theory.
We can restore the scaling property by using $\epsilon' = \epsilon/L$ as
a regularization parameter. Then we restore the original
scaling property which we express here in terms of the free energy $F$:
\begin{equation}
F(\epsilon',N,L) = \ln Z(\epsilon',N,L) = N \ln (L^2) + \ln z(\epsilon',N)
\;.
\end{equation}
Note that the reduced partition function $z(\epsilon',N)$ does no longer
depend on $L$. It is exactly this scaling property which will be important
later when we try to reconstruct the average density of instantons and
anti-instantons in the system.
We should emphasize that the price we have to pay for the
restoration of the scaling behavior is that the two limits
$\epsilon' \to 0$ and $L\to \infty$ cannot be interchanged, as with
growing L the effective cutoff $\epsilon = \epsilon'L$ becomes larger
than the average separation between zindons,
 $\langle R\rangle = 1/\sqrt{n}$. However,
when $\epsilon'$ becomes smaller the physical region where
$\epsilon = \epsilon'L < \langle R\rangle$
covers larger portions of the configuration space.
\newline
\newline
To study the behavior of the system numerically we first switch
off the interaction between instantons and anti-instantons i.e. we put
$\beta_1=0$. We then
have for $N_c=2$ the case of a simple Coulomb gas, where the
two different colors interact in the same way as the two different
electric charges. In the case of larger $N_c$ we have a more complicated
`multicolor' interaction which would then require a picture with
$N_c$ different kind of charges.
\newline
\newline
The variables of integrations for the theory
are the zindon coordinates $a_{Ai}$ and $b_{Ai}$. A direct numerical
calculation  of the partition function
(\ref{partfu1}) is technically not possible, because the
integrand has very sharp maxima: a small variation
of the zindon coordinates changes its value by orders of magnitude.
Instead of this one can apply a Metropolis algorithm: Starting with
a random configuration of zindons, step by step the position
of one zindon is varied by a small amount. If the new configuration
has a larger value of the weight $w_a w_b w_{ab}$ then this configuration
is accepted, if not then it is only accepted with the probability
$w_a w_b w_{ab}({\rm new})/w_a w_b w_{ab}({\rm old})$.
As Fig.~\ref{metropolis} shows, the plateau of important configurations
is reached only after 250.000 Metropolis sweeps for $N_c=3,4$. In the case
$N_c=2$ the number of steps necessary to reach the plateau depends on
the regularization parameter $\epsilon'$: It is large when $\epsilon'$
is small. In Fig.~\ref{metropolis} we have chosen ${\epsilon'}^2=0.001$.
Already after 20.000 steps one is very close to the plateau here.
It is now interesting to make a snapshot of
 the distribution of the zindons at the
plateau region. In Fig.~\ref{snapshot} we show snapshots of the zindon
ensemble at the plateau region after 500.000 Metropolis steps with
$\beta_1=0$. Each symbol represents zindons of one color. One
observes clearly, that for $N_c=2$ zindons of
different colors tend to condense into neutral pairs that can
be very densely packed. This is the reason for the fact that the
partition function is divergent and needs to be regularized in the
way described above. Such a condensation does not take place for
higher colors $N_c=3,4$.\newline\newline
We can put this in a more quantitative way by plotting the free energy
$F= \ln Z$ as a function of the regularization parameter $\epsilon$,
see Fig.~\ref{divergence}. As it should be expected by simple power counting
the free energy for the case $N_c=2$ diverges proportional to
$\ln(\ln(1/ \epsilon))$, while it is finite in the case of
higher $N_c$. In principle such a divergence does not pose
a real problem since physical quantities are only related to
logarithmic derivatives of the partition function, but it does
affect the effectiveness of the numerical studies considerably.
In fact, with the computer powers used, it was not
possible to go to ${\epsilon'}^2$ smaller than 0.001,
which is not very small. The smaller ${\epsilon'}^2$ becomes the larger
are the fluctuations in the partition functions
 until a point is reached where we do not get any
equilibrium within a reasonable time at all. We have to conclude from
this that the $N_c=2$ case is not a good system to be studied numerically.
\section{The stability of the instanton--anti-instanton ensemble}
\label{sec6}
\subsection{The case $N_c=2$}
\begin{figure}
\centerline{\psfig{figure=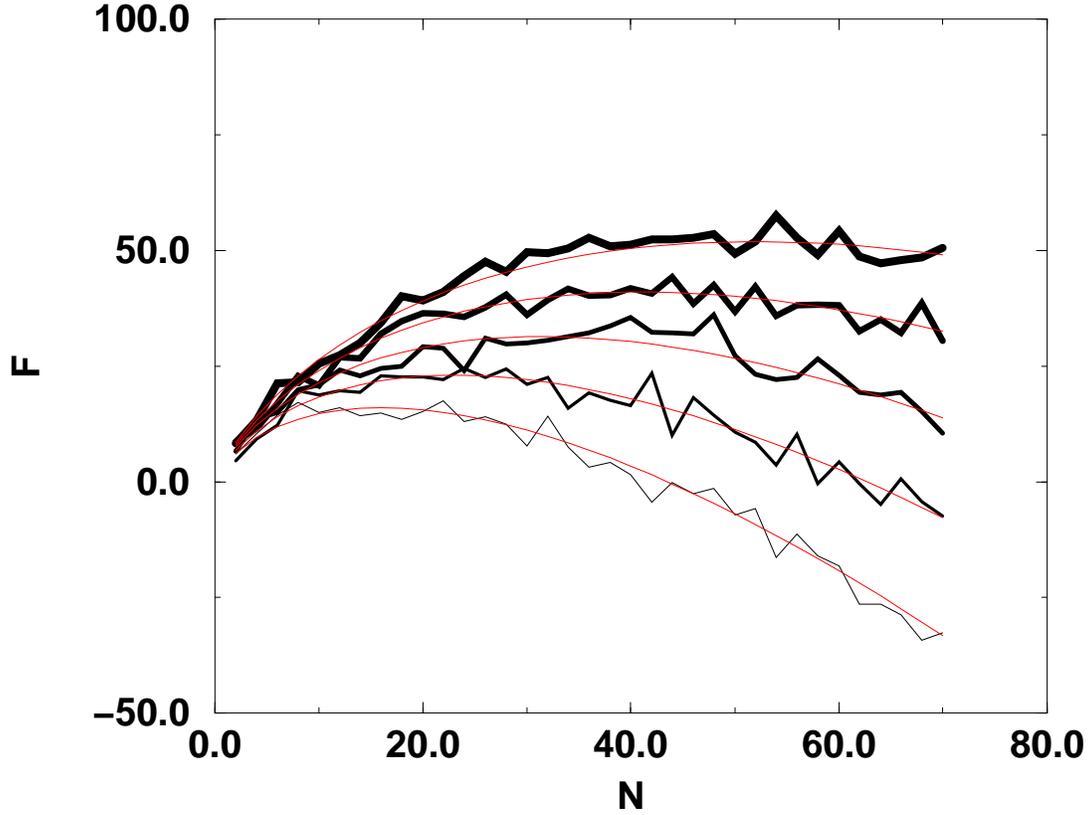,width=16cm}}
\caption{Density of the Coulomb gas ($N_c=2$): The curves in the figure
show the free energy $F$ versus the number of instantons $N$
using 20.000 Metropolis steps until equilibrium is reached. From
thin to bold we use the lengths
$L=0.5$,
$L=0.6$,
$L=0.7$,
$L=0.8$, and
$L=0.9$. The smooth solid curves show the combined  fit with
the scaling function $F(L,N) = N[\ln(L^2)-(\ln(N)-1-\ln(n_+))]+a$,
with a single $a$ and $n_+$ given to fit all five curves
simultaneously.
If we fit each curve separately
we get from thin to bold the following values for the density
$n_+=  62.341  $,
$n_+=  64.526  $,
$n_+=  66.027  $,
$n_+=  63.857  $, and
$n_+=  63.836  $.
In case of the combined fit we get for the density
the value $n_{+,{\rm best}} =  64.117  \pm  1.193     $, where the error is
the standard deviation derived from the five separate measurements
of $n_+$ given above. The cutoff parameter in this figure
is ${\epsilon'}^2 = 0.001$. The defect $\alpha$ is 0.481.}
\label{coul6}
\end{figure}

\begin{figure}
\centerline{\psfig{figure=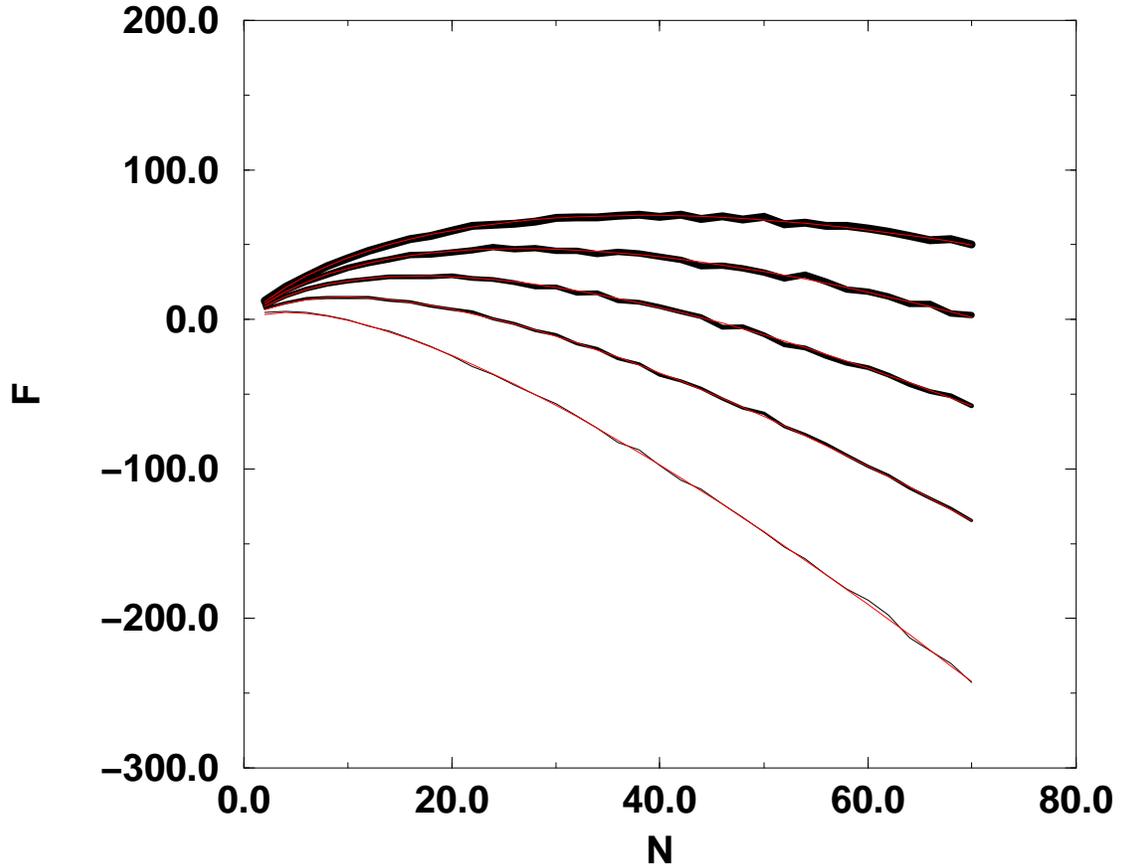,width=16cm}}
\caption{Check of the
Debye-H\"uckel approximation to the Coulomb gas ($N_c=2$).
The solid lines from
thin to bold give the measured value for a Coulomb gas with $\beta=0.1$
and lengths $L=2,L=3,L=4,L=5,L=6$. The thin smooth solid lines show the
prediction of the scaling law, which is F(N,L) =
$(2-\beta)N[ \ln(L^2)-(\ln(N)-1-\ln(n_+))]+a$, where $N$  in this
case is the number of instantons only, as we do not consider any
anti-instantons here.  The four curves
yield the value for $n_+= 1.06787\pm 0.0024$, which is in agreement
with the prediction of the Debye-H\"uckel approximation, which gives
$n_{+,\rm Debye}=1.0370$.}
\label{debye}
\end{figure}

\begin{figure}
\centerline{\psfig{figure=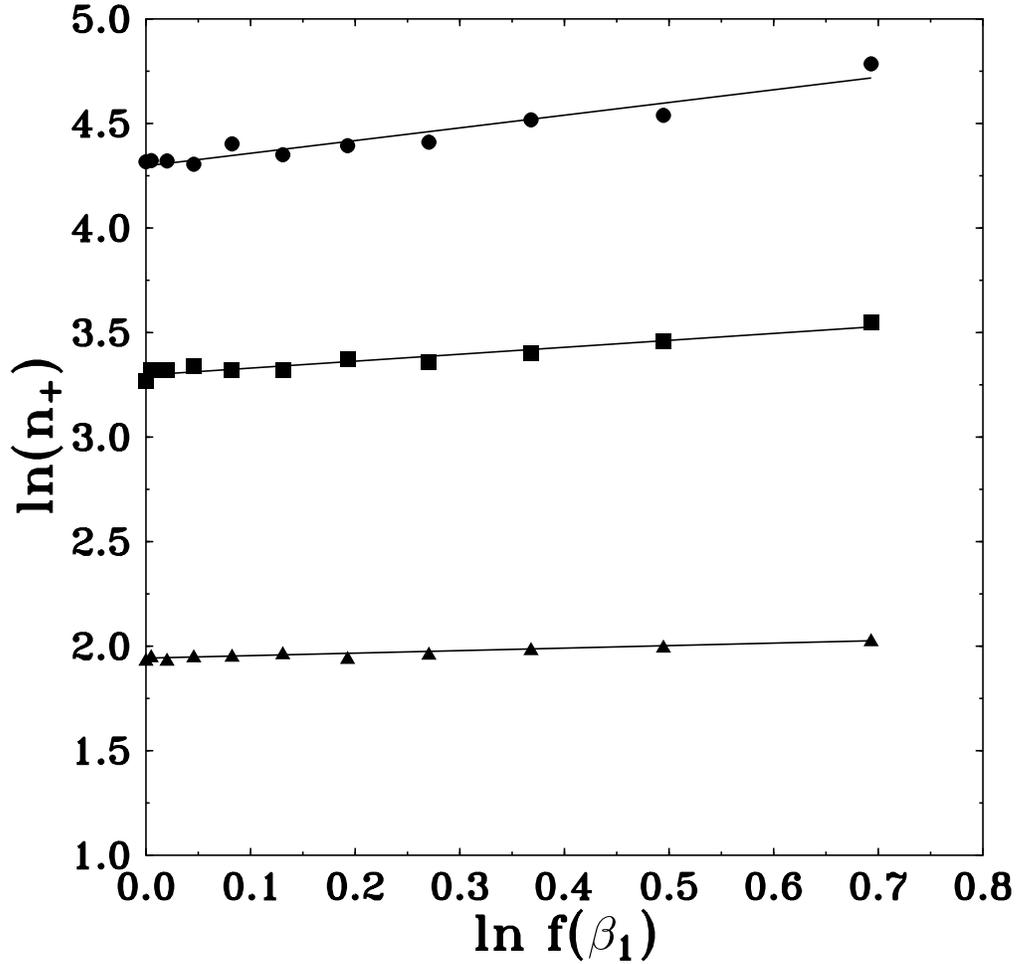,width=16cm}}
\label{running}
\caption{Fit of the equilibrium density to the form given by the
Debye-H\"uckel approximation. According to the Debye-H\"uckel 
approximation the logarithm of the
density $\ln n_+$ is linear in $\ln f(\beta_1)$, where $\ln f(\beta_1)=
\frac{1}{2}[ (1+\beta_1)\ln(1+\beta_1)+(1-\beta_1)\ln(1-\beta_1)]$.
In the figure the data for $\ln n_+$ are shown versus $\ln f(\beta_1)$,
where the triangles belong to the sample with ${\epsilon'}^2= 0.1$, the
squares to the sample with ${\epsilon'}^2= 0.01$ and the circles to
the sample with ${\epsilon'}^2= 0.001$. The solid lines are linear fits.
The corresponding fit constants are given in Tab.~\ref{debfit}.}
\label{runcoufig}
\end{figure}

The first question is whether this system
has a thermodynamic equilibrium, i.e. whether its density does not
change with the system volume $L^2$, when $L$ is large.
We start again with
the Coulomb gas of two noninteracting ensembles of instantons and
anti-instantons, i.e. $(\beta_1 = 0)$ and $N_c=2$.
Here it is  sufficient
to consider the instantons alone. In this case we simply have $N=N_+$ and
we plot the logarithm of the partition function, i.e. the free energy
as a function of the number of instantons $N$. The ideal situation
would be now that the number $N_{\rm max}$, where the partition function
has a maximum, is proportional to $L^2$. Then one could continue this
situation to an infinite instanton ensemble with the same density.
In Fig.~\ref{coul6} we show the partition function for the
Coulomb gas as a function of the number of instantons $N$ for various
box lengths $L$. The  simulation is done with the UV cutoff
 $\epsilon'^2=0.001$.
One sees that the free
energy is plagued by large fluctuations that correspond to fluctuations
in orders of magnitude in the partition function. A direct determination
of the maximum of the free energy is not reliable. Instead, one
can exploit the expected scaling behavior. If the system has a stable
density then the free energy must be of the form:
\begin{equation}
F(\epsilon',N,L) = \ln Z(\epsilon',N,L) =
N\{\ln (L^2) -[ \ln(N) -1 -\ln(n_+(\epsilon'))]\} + a \;,
\label{fitform}
\end{equation}
with the two fit constants $a$ and $n_+(\epsilon')$, and
$\epsilon'=\epsilon/L$ being fixed in order to preserve the scaling
behavior of the initially unregularized theory. \Eq{fitform}
is obtained in the following way: From the inspection of \eq{partfu1}
it is clear that the free energy depends on $L$ via $N\ln(L^2)$. Requiring
now that the number $N$ at the maximum of the free energy $F$ is proportional
to $L^2$ leads to a differential equation
with the solution given by \eq{fitform}. \Eq{fitform}
has the advantage that $n_+(\epsilon')$ gives immediately
the density of the system. The symbol $n_+$ refers to
the instanton density.
Unfortunately, the system does not exhibit a stable density. The reason
is, again, the tendency of the system to condense into neutral
pairs corresponding to small-size instantons. When it happens, it is not
reasonable anymore to write down the particle-identity factor $1/(N!)^2$
in the partition function: it should be rather replaced by the first power
of the factorial. Therefore, we allow for an additional
`conbinatorical defect' term in the fit, which reads $\alpha \ln N!$, so that the full
fit function has now the form:
\begin{equation}
F_{\rm fit} (\epsilon',N,L) =
N\{\ln (L^2) - [ \ln(N) -1 -\ln(n_+(\epsilon'))]\}
+ \alpha \ln N!+ a \;.
\label{fitform2}
\end{equation}
The smooth solid lines in Fig.~\ref{coul6} are the fitted curves, which
fit quite well to the heavily fluctuating free energy. One should expect
that the defect becomes smaller when $\epsilon'$ becomes smaller.
As shown in Tab.~\ref{tab1}, one sees indeed such a behavior.
\newline
\newline
Now arises the question how reliable our simulations are, because
so far we did not find strictly speaking a stable system. The situation
 is also not very comfortable because in the limit
$\epsilon'=0$, where a stable phase should exist the density $n_+$
is divergent for $N_c=2$. We therefore study now a different regularization
method:
It consists in introducing a
temperature $T=1/\beta$ in the zindon interactions, with $\beta<1$.
We reach the physical limit by taking  $\beta \rightarrow 1$. For small
values of $\beta$ we expect
behavior of the system to be consistent with the predictions
of the Debye-H\"uckel approximation. In Fig.~\ref{debye}
we show the free energy
for $\beta=0.1$ versus the number of instantons $N$.
The solid lines are the fits using the form:
\begin{equation}
F_{\rm fit,\; Debye}(N,L) =
N\{\ln (L^2) - [ \ln(N) -1 -\ln(n_{+,{\rm Debye}})]\} + a \;.
\label{fitformdeb}
\end{equation}
We find a stable density for various box lengths $L$.
The five curves yield a value for $n_+= 1.06787\pm 0.0024$,
which is in agreement with the prediction of the Debye-H\"uckel approximation:
(\ref{densdeb}) gives $n_{+,{\rm Debye}}=1.0370$.
\newline
\newline
We now switch on the interaction between instantons and anti-instantons
putting $\beta_1>0$.
Again we can only determine a kind of
stable density $n_+(\epsilon')$, if we allow for a defect
parameter. Tab.~\ref{runcou} shows the results for $n_+(\epsilon')$
for three different values of $\epsilon'$ as a function of $\beta_1$.
In case of $\beta_1=0$ we can compare the density to the one in
Tab.~\ref{tab1}, both are results of independent simulations.
A comparison of the upper line in Tab.~\ref{runcou} with
corresponding entries in  Tab.~\ref{tab1} gives an idea of the statistical
error; it is not enormous, but certainly not negligible.
\newline
\newline
We observe that the density rises only slowly with
$\beta_1$. This is in agreement with the results of the Debye-H\"uckel
approximation,
which also predicts a slow rising of the density with $\beta_1$.
For $N_c=2$, the equilibrium density is given by (see \eq{densdeb}):
\begin{equation}
n_+ =
\frac{1}{2}\Bigg[ f(\beta_1)  \pi \beta
e^{2\gamma_E}2^{\frac{2}{\beta}}\Bigg]
^{\frac{\beta}{2-\beta}};\quad
f(\beta_1) = (1+\beta_1)^\frac{1+\beta_1}{2}(1-\beta_1)^\frac{1-\beta_1}{2}
\;.
\label{debpred}
\end{equation}
The function $f(\beta_1)$ varies between 1 and 2 for $\beta_1$
varying between 0 and 1.
For the real $CP^1$ model, when $\beta =1$, we can expect at least a behavior
linear in $\ln f(\beta_1)$ for the logarithm of the density $\ln n_+$.
Fig.~\ref{runcoufig} shows, indeed, that we can obtain a good
linear fit, using this assumption. The numerical results for the linear
fit can be read off from Tab.~\ref{debfit}.
The growth of the intercept $a$ when ${\epsilon'}^2$ goes to
zero reflects the fact that in this theory the density is divergent.
The growth of the slope $b$ is also natural, since at
$\epsilon' \rightarrow 0$ we expect $b=1$ from \eq{debpred}.
In general one sees that the effect of the instanton--anti-instanton
coupling on the density is comparatively small as predicted also
by the Debye-H\"uckel approximation.
\newline
\newline
Finally, let us discuss the running of the  instanton--anti-instanton
coupling $\beta_1$. From its derivation in Sec.~\ref{sec3.2} it is
seen that $\beta_1=\pi/g^2$ ( at $N_c=2$)
is related to the bare coupling constant of the theory $g^2$. However,
physical quantities cannot depend on the bare couplings, but only on the
renormalized ones. It means that by considering quantum corrections
to the classical instanton--anti-instanton interactions one would be able
to observe that the bare coupling constant $g^2$ (or $\beta_1$) is replaced by
the running coupling depending on the characteristic scale in the
problem at hand. In our case the only dimensional scale is the average
separation between instantons and anti-instantons or, else, the
density of zindons, $n$. From  renormalization-group arguments one can
write the 1-loop formula
\begin{equation}
\beta_1 = \frac{1}{2} \left( \ln \frac{N}{L^2} +C\right)\;,
\label{runc}
\end{equation}
however, the constant $C$ remains unknown. In order to determine $C$
one has to perform the renormalization of the instanton--anti-instantons
interaction with a very high precision; this has not been done.\
\newline
\newline
Keeping in mind that the dependence of the instanton density on $\beta_1$
is very mild in the whole range $0<\beta_1<1$ (see Fig.~\ref{runcoufig} and
Tab.~\ref{runcou}) and that the dependence of $\beta_1$ on the
density is also quite weak (see (\ref{runc})), and given the theoretical
uncertainty in the constant $C$, we cannot reliably determine the density
self-consistently by taking into account the running of $\beta_1$.
In the case of  $N_c=2$ by far a more important factor
affecting the density is the necessary ultraviolet cutoff $\epsilon$.
\subsection{The case $N_c=3,4$}
\begin{figure}
\centerline{\psfig{figure=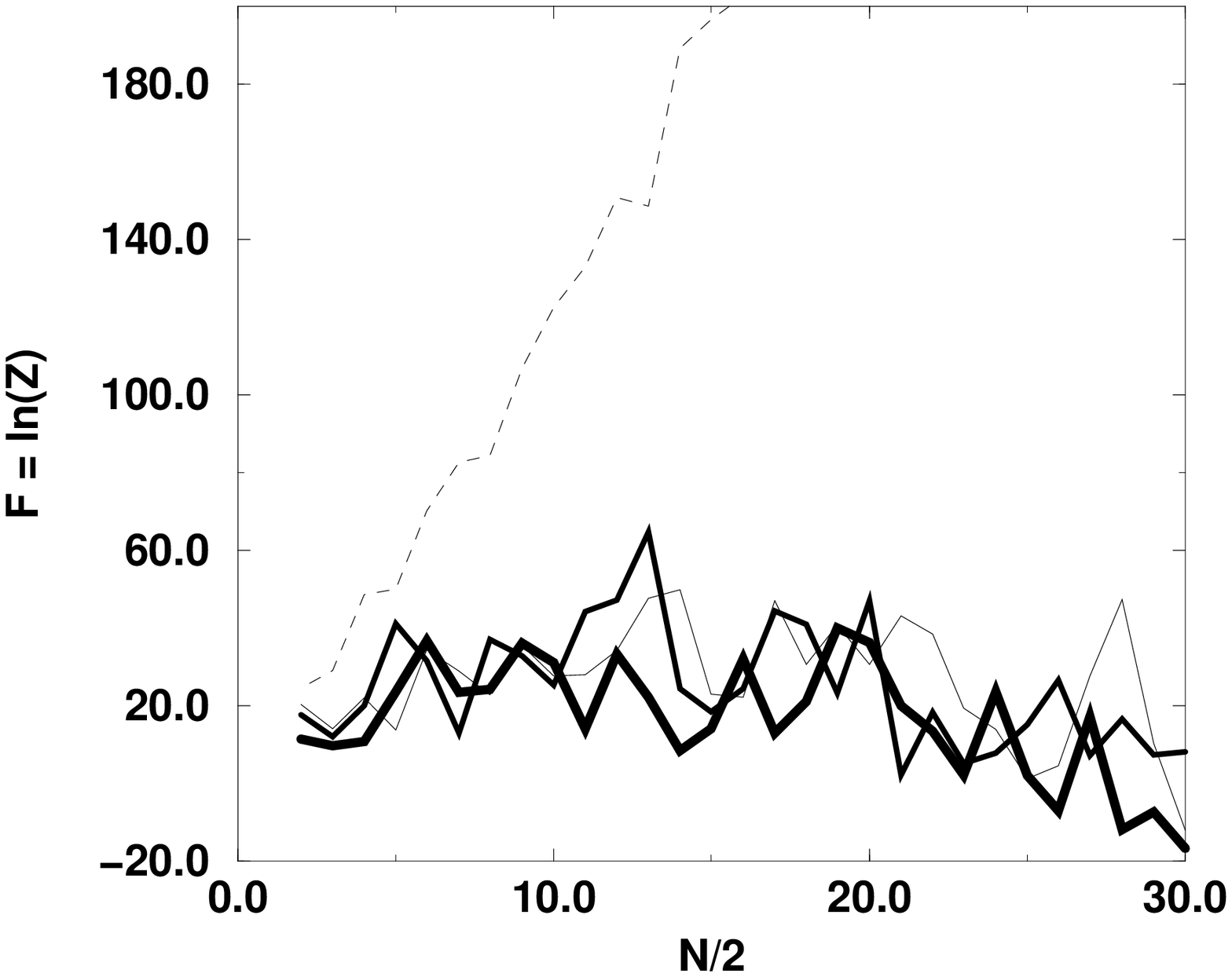,width=8cm}
            \psfig{figure=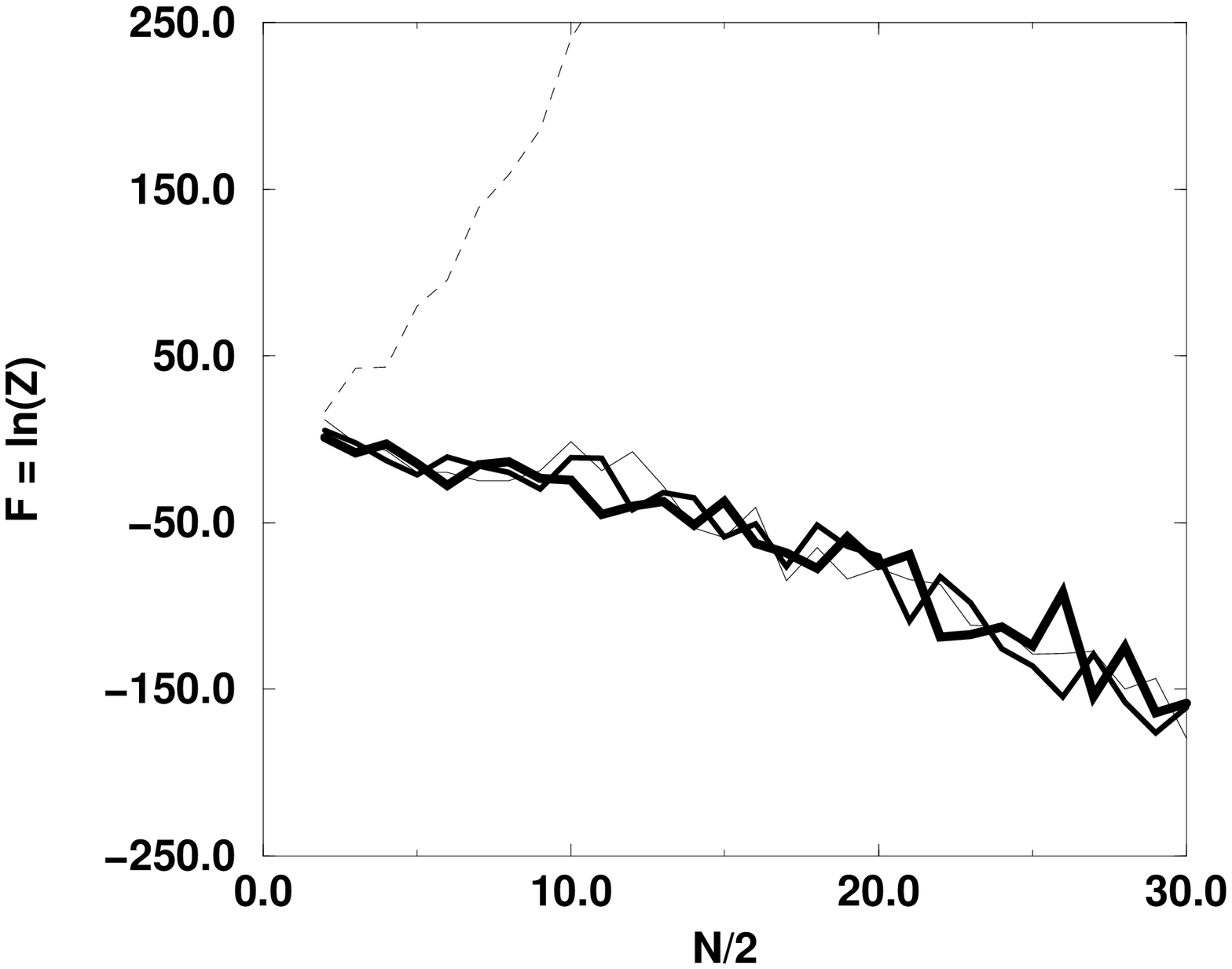,width=8cm}}
\caption{Free energy versus the number of instantons (equal to
the number of anti-instantons), $N/2=N_-=N_+$. The figures show the
free energy versus $N/2$ for $N_c=3$ (on the left) and $N_c=4$
(on the right). The solid lines from thin to bold are for
$\beta_1=0.0$,
$\beta_1=0.2$, and
$\beta_1=0.4$.
The dashed line
is the free energy for the  value $\beta_1=1$, which shows
all signs of a collapse.}
\label{densities}

\end{figure}

\begin{figure}
\centerline{\psfig{figure=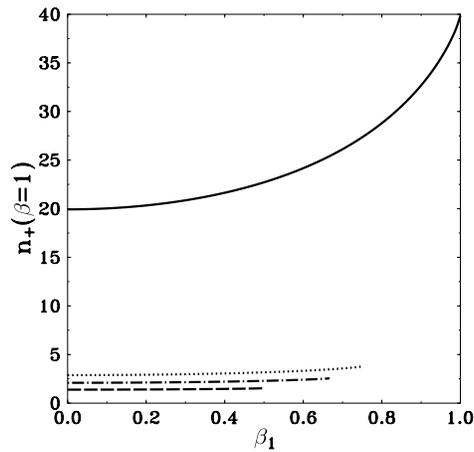,width=8cm}}
\caption{Prediction of the Debye-H\"uckel approximation for the
density at $\beta=1$. The solid line is the Debye-H\"uckel prediction
for $N_c=2$, the dotted line for $N_c=3$, the dashed-dotted line 
for $N_c=4$, and the dashed line shows
the limit $N_c\to \infty$. The lines end at critical couplings.}
\label{debyegraph}

\end{figure}

In the case $N_c=3,4$ we are not plagued with the necessity to introduce
a regulator. However, we are confronted now with enormous fluctuations.
In Fig.~\ref{densities} the free energy is plotted for several values
of $\beta_1$. Already in the free energy the fluctuations are of an order
of magnitude, so that a reasonable determination of the equilibrium
density is not possible. However, we could at least try to compare the
situation to the one in the Debye-H\"uckel approximation for $N_c=3,4$.
First of all it is necessary
to mention that our partition function (\ref{partfu1}) is not
based on two-body interactions only,
and is therefore not directly comparable to the one used to derive the
density (\ref{densdeb}). But, as we will see now, some of the features
are nevertheless strikingly in parallel. First of all it turns out
that for all values of $\beta_1$  not too close to the critical value, 
the free energy does
not change much (as compared to the extremely large fluctuations). This
is in parallel to the fact that the equilibrium density given by
the Debye-H\"uckel approximation grows only slowly with $\beta_1$, see
Fig.~\ref{debyegraph}. Another feature of the Debye-H\"uckel result
for the equilibrium density is that it is  comparatively large for $N_c=2$ 
and then falls rapidly to the asymptotic value
$n_+(N_c=\infty,\beta=1,\beta_1=0) =
\pi^{1/7}\exp(2\gamma_E/7) \approx 1.3888$. The second point, where there
is something in parallel between the Debye-H\"uckel
result and the simulations,
concerns the critical behavior. The  Debye-H\"uckel formula for the density
becomes meaningless when $\beta_1 \geq N_c/2(N_c-1)$. Hence for $N_c=3$
we find $\beta_{\rm crit}= 3/4$ and for $N_c=4$
we find $\beta_{\rm crit}= 2/3$. In the simulations it
is seen  (Fig.~\ref{densities})
that for values of $\beta_1\simeq 1$ the system behaves
abnormally and shows clear signs of a collapse.
\section{Size distribution}
\label{sec7}
\begin{figure}
\centerline{\psfig{figure=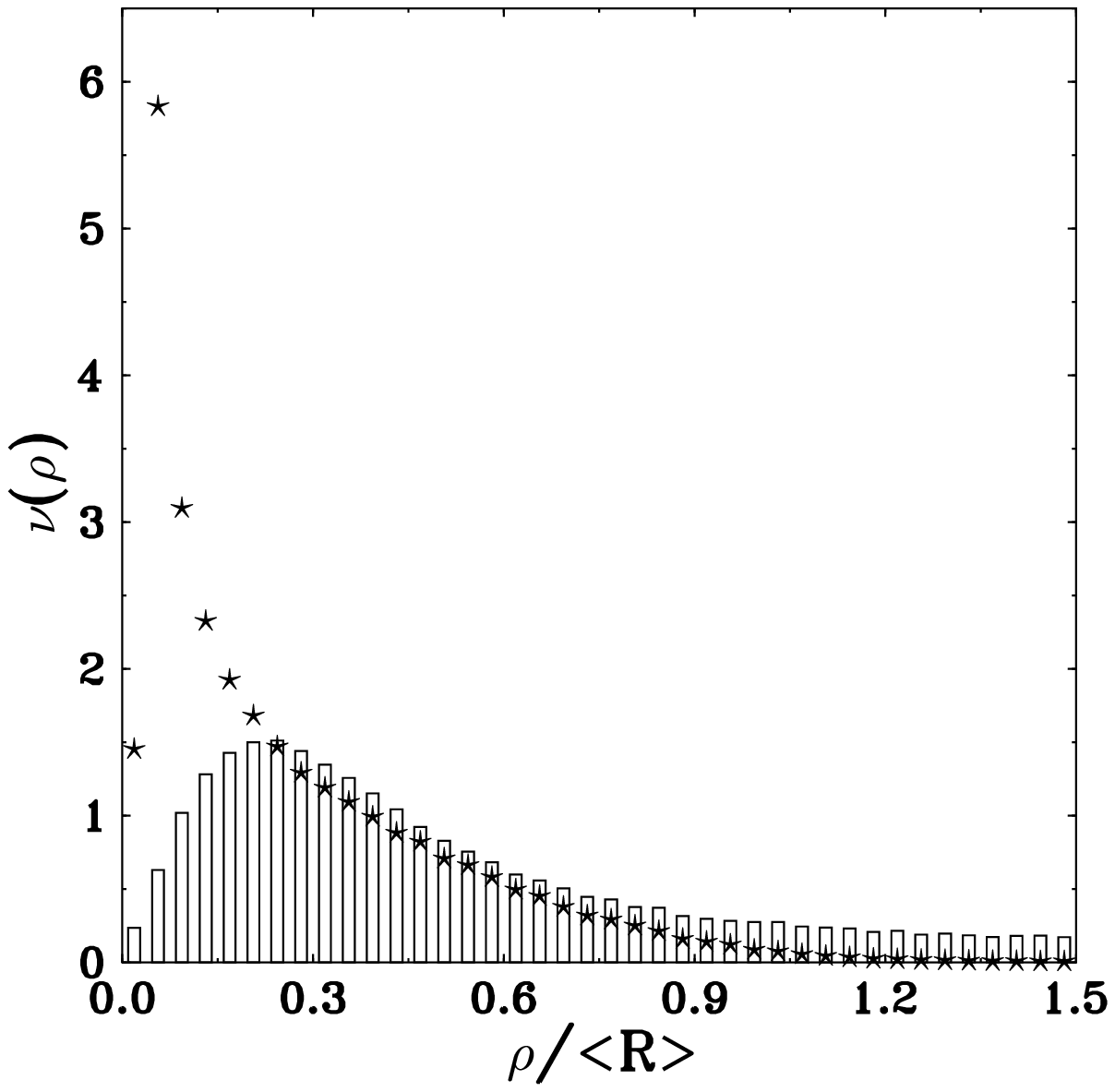,width=8cm}}
\centerline{\psfig{figure=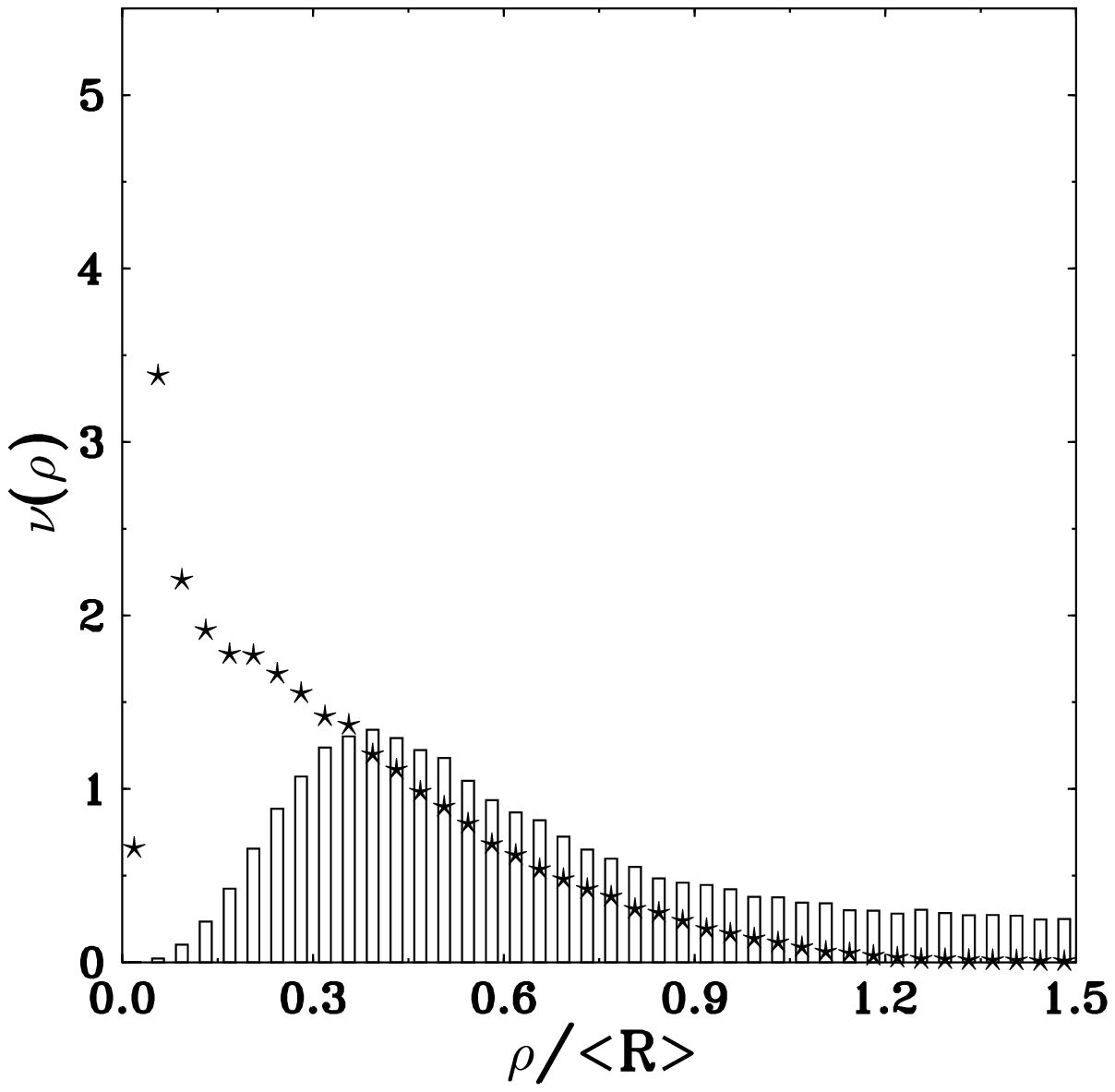,width=8cm}
            \psfig{figure=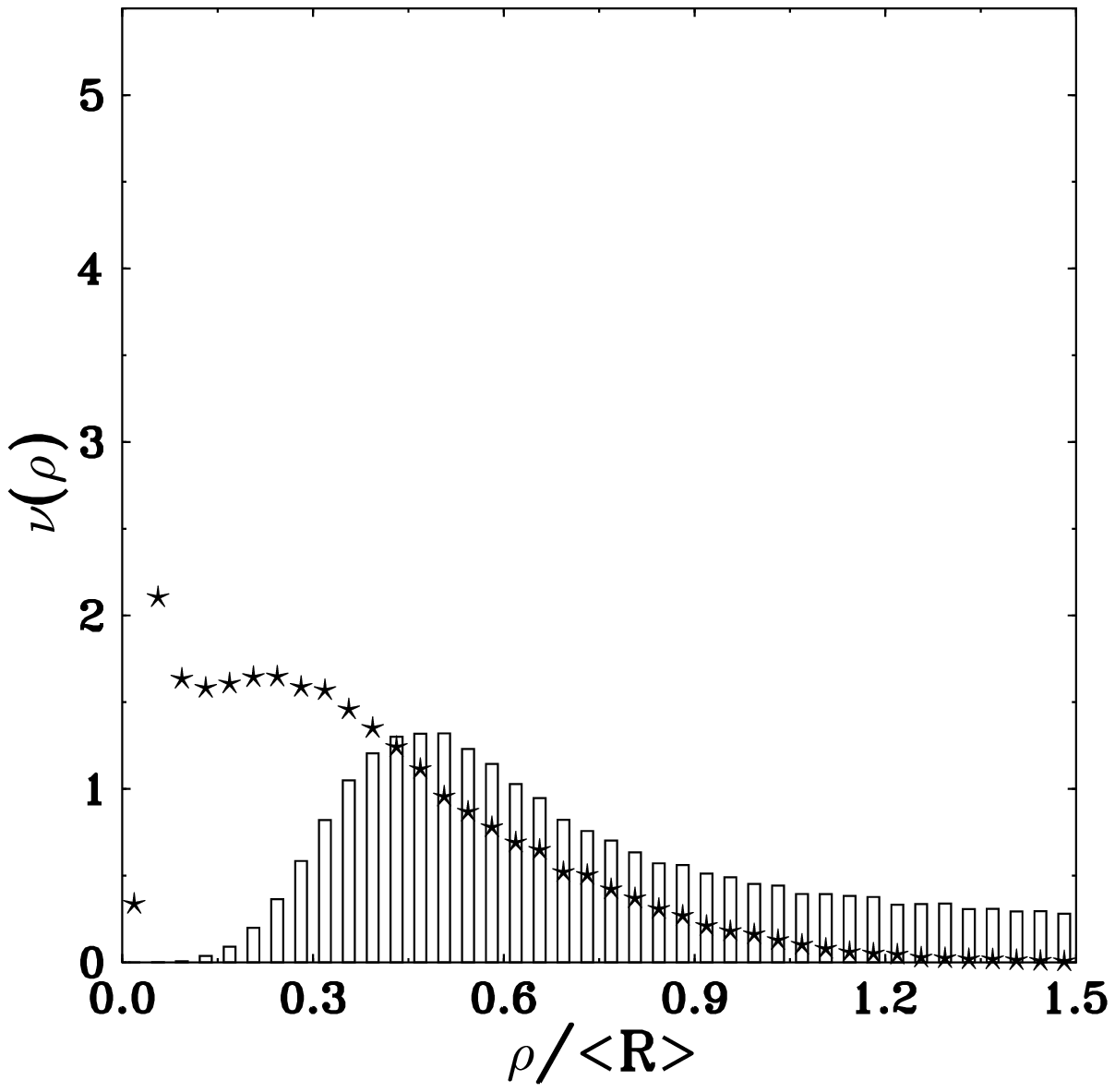,width=8cm}}
\caption{Instanton size distribution following from random (non-interacting)
zindons are displayed for  $N_c=2$ (on the top),
$N_c=3$ (bottom left), and $N_c=4$ (bottom right) for $N_+=N_-=8$.
The histograms show the size distributions `geometrically'
defined by grouping zindons according to
their dispersion $\rho$. The stars show the result for the `lattice' method
using a $100\times 100$ grid.}
\label{methodsflat}
\end{figure}

\begin{figure}
\centerline{\psfig{figure=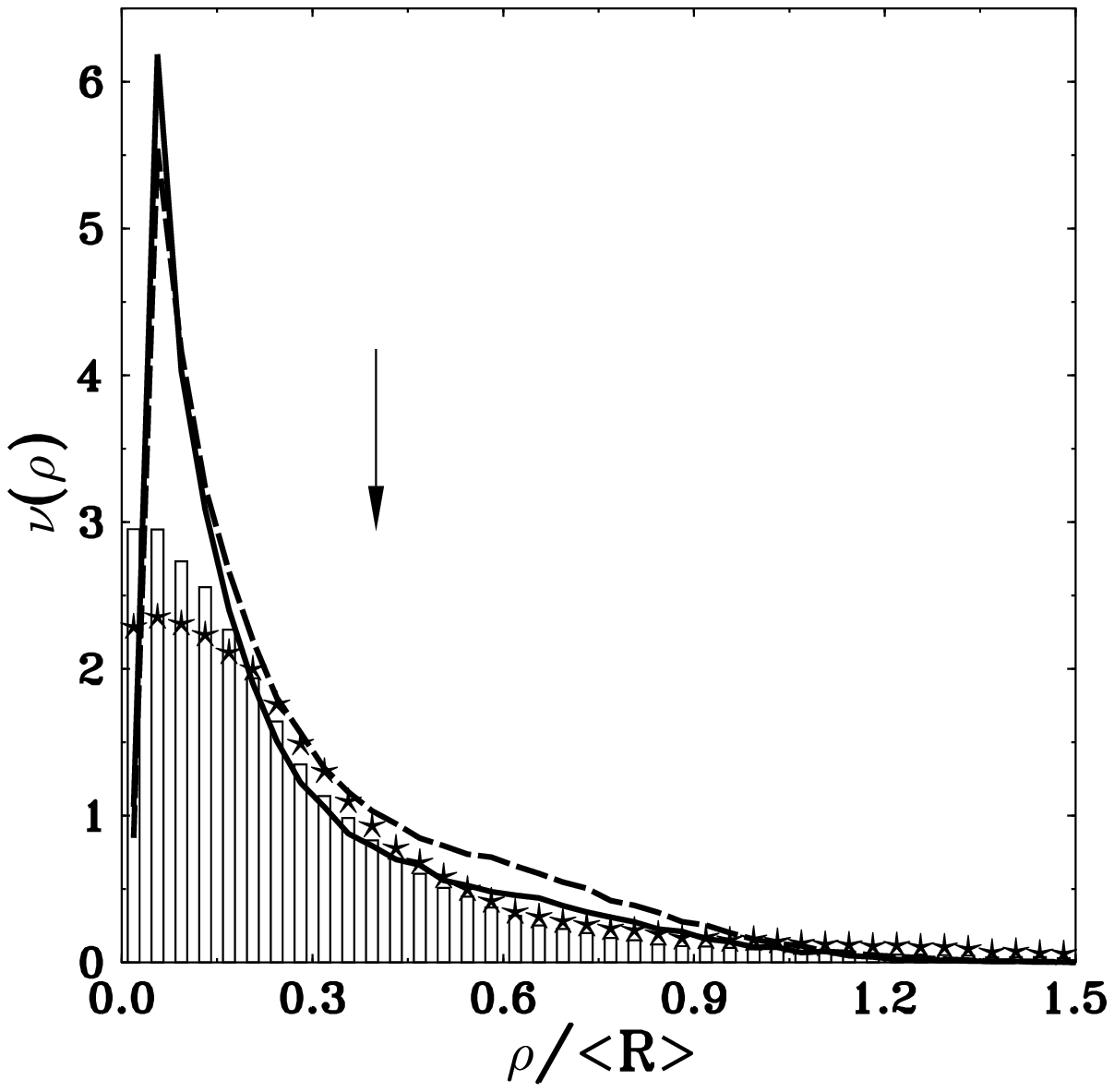,width=8cm}
            \psfig{figure=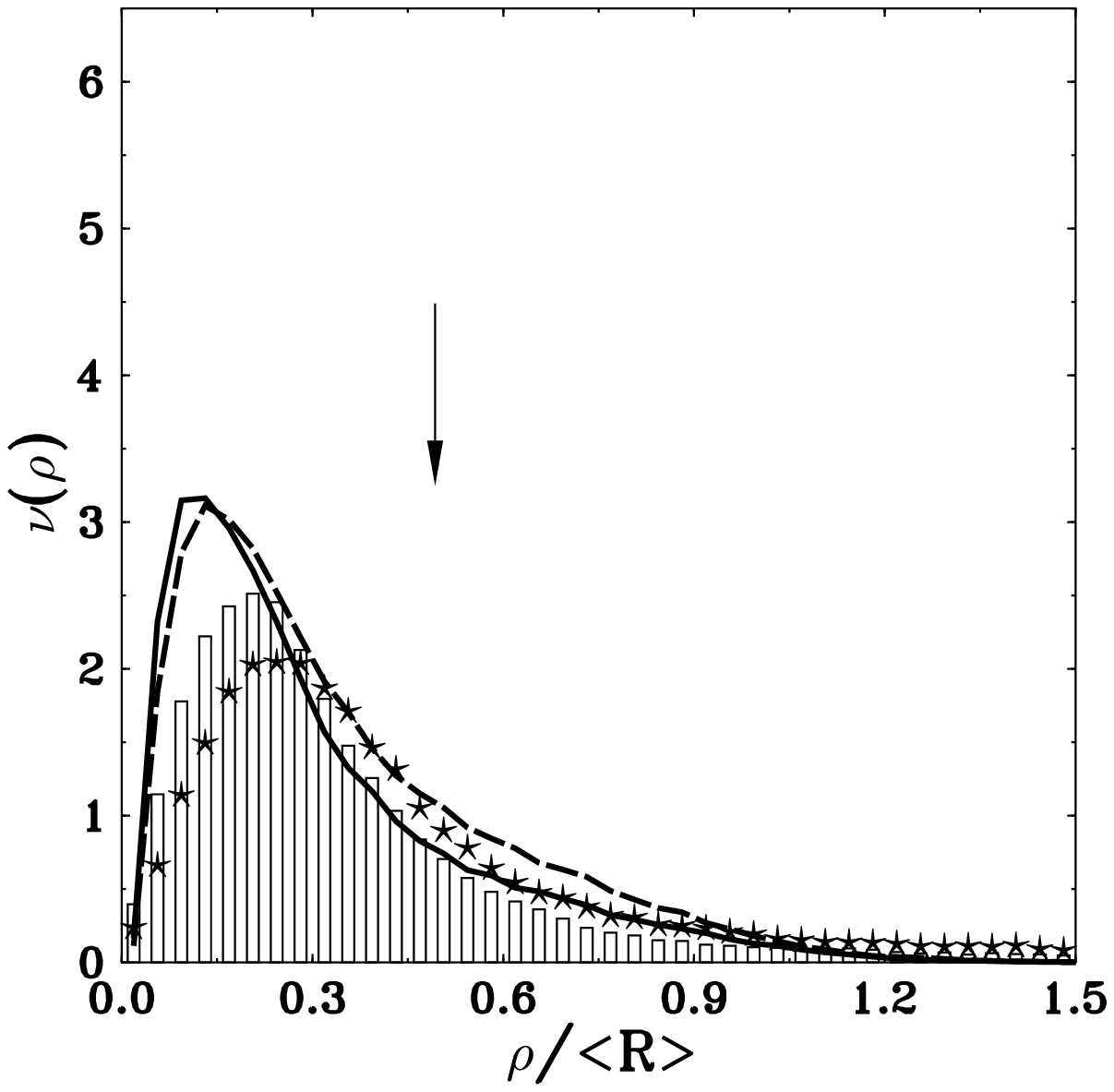,width=8cm}}
\caption{Instanton size distributions with zindon interactions switched on
are displayed for $N_c=3$ (left), and $N_c=4$ (right)
for $N_+=N_-=8$.
The histograms show the `geometric' size distributions, and
the solid lines the size distributions seen by the
`lattice' method, again using a $100\times100$ lattice.
The instanton--anti-instanton coupling constant
for the solid lines and the histograms is
$\beta_1=0.0$. The stars and the dashed lines show
the `geometric' and `lattice' size distributions for the case of
$\beta_1 =0.5$.
Instanton sizes are plotted in units of the average separation
$\langle R\rangle$.
The arrows show the maxima of the corresponding
`geometric' size distributions obtained in the non-interacting case of
Fig.~\ref{methodsflat}.}
\label{sizedist}
\end{figure}

A very important question is what is the size distribution
of instantons in the ensemble. Is the average size $\langle \rho \rangle$
smaller or larger than the average separation between instantons
$\langle R \rangle = n^{-1/2}$? In the former case
one can say that instantons are dilute while in the latter case one says that
instantons `melt', so that individual instantons have not much
sense.
\newline
\newline
We are now in a position to study this problem quantitatively.
The zindon parametrization of instantons is an ideal tool
for that since it allows for both extreme cases. If
zindons of different `color' are uniformly distributed in space
that would mean the instantons `melt'. If zindons of $N_c$ colors
tend to form well-isolated color-neutral clusters,
it means instantons are dilute.
The size distribution of instantons has been measured on
the lattice for $N_c=2$ \cite{MS94} after a cooling procedure has been
applied to remove short-range fluctuations. (For a recent comparison
of different cooling techniques on the $CP^{N_c-1}$ model see also
\cite{ACEP99}.)
A very delicate question in
connection to this is, how to identify  instantons on the lattice.
Several procedures are in use \cite{Ne98}. In this paper we want to
compare two ways of extracting the instanton content. The first one, which
we call the `geometric' method, makes use of the zindon picture.
From the sample of all zindons the color neutral $N_c$-plet which has
the least dispersion about the common center is chosen to be an instanton.
Then all zindons belonging to this instanton are removed from the sample,
and the procedure is iterated to find the next group of zindons that
has the smallest dispersion and so on, until the whole sample is
grouped into instantons. The second procedure, which we call
the `lattice' method as it is more close to the ones used in lattice
studies, is looking for local maxima
of the topological charge density $q(x)$ on a $N_{\rm grid}
\times N_{\rm grid}$ lattice. An instanton is assumed if a grid point
has a larger topological charge than its surrounding 8 nearest
neighbors.
In order to improve the accuracy in finding the center points and the sizes,
we construct an interpolating polynomial,
\begin{equation}
q(x,y) = ax^2y^2+bx^2y+cxy^2+dx^2+ey^2+fxy+gx+hy+j,
\end{equation}
calculated from the 9 grid points consisting
of the grid point under consideration and its 8 neighbors. From this
interpolating function we find coordinates of the local extrema $x_E$.
An extremum is accepted if it lies
inside the $3\times 3$ grid used for the interpolation.
In order to classify whether we have found a
local maximum, minimum or a saddle point we construct the
matrix of second derivatives at the extremum $x_E$:
\begin{equation}
\left(
\begin{array}{cc}
      \left(\frac{\partial}{\partial x}\right)^2  &
      \frac{\partial}{\partial x}\frac{\partial}{\partial y} \\
      \frac{\partial}{\partial x}\frac{\partial}{\partial y} &
      \left(\frac{\partial}{\partial y}\right)^2\\
\end{array}
\right)q(x_E) \;.
\end{equation}
If the two eigenvalues $\lambda_1$ and $\lambda_2$ of this matrix
have a different sign then $x_E$ is a saddle point and is rejected. For
negative  $\lambda_1$ and $\lambda_2$ we have a local maximum and
respectively a local minimum if both $\lambda_1$ and $\lambda_2$
are positive. The two eigenvalues give at the same time a measure for
the size $\rho_i^2 = \sqrt{8/|2\pi \lambda_i|}, \;i = 1,2$ (they
should coincide for an ideal spherically-symmetric instanton).
This size can be
compared to the one derived from the value of the topological charge
at the maximum $\rho_3^2 = 2/|2 \pi q(x_E)|$. Only if all three sizes
are smaller than $L/2$ is the instantons or anti-instanton accepted and
the size is then taken to be the geometrical mean of the three
definitions,
$\rho = (\rho_1 \rho_2 \rho_3)^{1/3}$.
It should be noted that usually $\rho_{1,2,3}$ are not drastically different.
We should also stress that a reliable identification of all local
extrema is achieved only with a fine grid. With a 100$\times$100 grid
we find almost the same number of maxima (instantons) and minima
(anti-instantons) as we put in `by hands' by fixing the number of
zindons and anti-zindons.
\newline
\newline
We first test both methods, the `geometric' and the `lattice',
in a toy model where the positions of zindons and anti-zindons are
random: the interactions are switched off completely ($\beta=0$ and
$\beta_1=0$) so that the distribution of zindons in space is given
by a flat measure $\prod d^2a_{Ai}\prod d^2b_{Aj}$. We generate
configurations of zindons with this measure and compute the
topological charge density $q(x)$ of a given configuration from
the integrand of \eq{acttop}. Instantons and anti-instantons
are then identified by the two methods described above.
Taking many configurations we compute the distribution of the
instanton sizes, see Fig.~\ref{methodsflat}.
\newline
\newline
For the flat measure the small-$\rho$ side of the distribution can be
immediatelly evaluated from dimensions to be $\nu(\rho)\sim
\rho^{2N_c-3}$. The geometric distribution (shown by histograms) follows
this prediction whereas the lattice method strongly deviates from it.
[One can mimic the smooth geometric distribution by the lattice method
by taking a crude grid but then many instantons are lost
as they ``fall through'' the grid.] The fact is that the topological
density computed from randomly distributed zindons shows many sharp
and narrow peaks. The larger number of instantons we take
(at fixed density) the more sharp the peaks become: it clearly indicates
a non-thermodynamic behavior of the system.
\newline
\newline
This pathological behavior is due to the utmost instability of the
product Ansatz \ur{Ansatz} in the case when the interactions are swithched
off. Indeed, the sizes of instantons are determined by \eq{size} where
the role of the weights $c_A$ is played by long products \ur{cA}. These
products are extremely unstable: sufficient to change the positions
of very distant zindons, and $c_A$'s will change. With zindons distributed
according to the flat measure $c_A$'s fluctuate by orders of magnitude,
and the more zindons are taken the stronger they fluctuate. It means
that there will always be a zindon of a certain color whose weight
$c_A$ is much much larger than those of the other zindons forming
an instanton. According to \eqs{center}{size} the extremum of the
topological density will then coincide with the position of the zindon
of that particular color, and the size of the corresponding instanton
will be close to zero. This is why we see too many small-size
instantons by the lattice method. See also the discussion around
Fig.~\ref{illustration} below.
\newline
\newline
When we switch on the interactions zindons get correlated and the
product Ansatz should not be pathological any more. In particular,
from the experience of the Debye--H\"uckel approximation we know
that the correlation functions of zindons get exponentially suppressed
at large separations. Therefore, we expect that the topological density
will not have unphysical sharp peaks at small $\rho$ as in the case of the
flat measure, and the two methods of instanton identification should
come closer. In practice, however, one cannot be sure beforehand
that the thermodynamic equilibrium is fully reached, given the limited
computer time for configuration generation.
\newline
\newline
We switch on the interactions, first, only of zindons belonging
to same-kind instantons ($\beta_1=0$)
and then adding interactions
between instantons and anti-instantons ($\beta_1=0.5$),
see Fig.~\ref{sizedist}.
We show only the cases $N_c=3,4$ as for $N_c=2$ the size distribution is
cutoff dependent. For the `lattice' method we use again the
$100\times100$ grid.
Unfortunately, the fluctuations of the partition function
in the simulations are so large that we have to apply a trick
in order to get meaningful results. We let the Metropolis algorithm
work until a plateau is
reached. Then we make many measurements of the size distribution
at the plateau, all weighted with a weight averaged over the plateau.
We repeat this for several starts but we essentially see the size
distribution for the plateau possessing the largest free energy.
\newline
\newline
The price we have to pay for this somewhat truncated simulation
is that the unphysical peaks at small $\rho$, as determined
from the lattice method, are still present though they are not so awkward
as in the non-interacting case. As $N_c$ increases the small-$\rho$
peak shrinks, therefore one can speculate that at large $N_c$ it is
easier to reach the thermodynamic equilibrium. Except for very small $\rho$
the two methods of instanton identification seem to produce rather
similar distributions.
\newline
\newline
The geometric size distribution exhibits the expected behavior
at all $\rho$. In particular, at small $\rho$ it behaves as
$\nu(\rho) \sim \rho^{b-3}$ where $b=N_c$ is the 1-loop Gell-Mann--Low
coefficient, which is to be expected from dimensional analysis.
\newline
\newline
We observe that the instanton--anti-instanton interaction
does not influence the size distributions significantly. This is seen
both from geometric and lattice methods by comparing
curves in Fig.~\ref{sizedist} with $\beta_1=0$ and $\beta_1=0.5$.
The case $\beta_1=0.5$ has a slightly broader distribution.
Qualitatively, it should be anticipated since the interactions between
zindons and anti-zindons tend to `pull out' individual zindons
out of their neutral clusters, however, the effect is small.
\newline
\newline
As to the interactions of the same-kind zindons, it clearly leads
to smaller instanton sizes; this could be also anticipated.
Indeed, the zindon interactions are such that same-color zindons
are repulsive while different-color zindons are, on the
average, attractive. Such interactions lead to preferred
clustering of zindons into color-neutral groups.
(This is especially clear for $N_c=2$ where the
Coulomb interaction leads, as we have seen, to a collapse of
zindons into neutral pairs as one removes the ultraviolet
cutoff, $\epsilon\rightarrow 0$).
In Fig.~\ref{sizedist} we show by arrows at the top of the plots the
positions of the `geometric' distributions maxima in the non-interacting
case of Fig.~\ref{methodsflat}:
both  for $N_c=3$ and 4 the maxima are noticeably shifted to
smaller sizes when one switches on the interactions.
The `lattice' method shows this shift only in the case $N_c=4$
where the unphysical small-size peaks of the topological density
are suppressed.
\newline
\newline
In Fig.~\ref{illustration} we plot the topological
density obtained from a typical
configuration of zindons at $N_c=3$ with all interactions taken into
account. We see that the peaks of the topological density in general
correlate with the positions of the centers of instantons found by the
`geometric' method (shown by triangles in Fig.~\ref{illustration}).
In some cases, however, the peaks of the topological charge
density sit not in the centers of the 3-plet of zindons forming an instanton
but rather near one of the three zindons. This is the consequence
of the instability of the weight coefficients $c_A$, discussed above,
and is the actual cause of the small-$\rho$ peaks in the `lattice'
size distributions.
\newline
\newline
Finally, we would like to draw attention to the fact that
in Fig.~\ref{sizedist} we have plotted the instanton sizes in units
of the average separation between instantons, defined as
$\langle R\rangle = (N/V)^{-1/2}$. It is remarkable that both for $N_c=3$ 
and 4 and independently of the method used to identify instantons, 
the average instanton size is definitely smaller than the average separation.
It means that instantons formed of $N_c$-plets of zindons are,
on the average, dilute. Though one can always find a huge instanton
overlapping with many others, the probability of such an event is relatively
low.
\section{Conclusions}
We have formulated the statistical mechanics
of instantons and anti-instantons in the $d=2$ $CP^{N_c-1}$ models
in terms of their `constituents' which we call `zindons'.
We have derived the interactions of same-kind and opposite-kind
zindons for arbitrary $N_c$. At $N_c=2$ they come to logarithmic Coulomb
interactions known previously. At $N_c>2$ the interactions are of
a many-body type though one can reasonably approximate them by
two body logarithmic interactions with charges depending on the `color' of
zindons.
\newline
\newline
At high temperatures the system can be studied analytically by ways
of the Debye-H\"uckel approximation, at any $N_c$.
The physical case corresponds to the temperature $\beta=1$ and can be studied
by a Metropolis-type simulation. The numerical study shows that the
Debye-H\"uckel approximation is, as expected, accurate at small $\beta$, but
is qualitatively valid even at $\beta=1$. Both analytical an numerical
studies indicate that the effect of instanton--anti-instanton interactions
(in our language the zindon--anti-zindon interactions)
is not significant: the interaction of same-kind zindons or,
else, the multi-instanton measure, is by far more important for
the statistical mechanics of the instantons--anti-instanton ensemble.
\newline
\newline
At $N_c=2$ the system is, strictly speaking, collapsing into instantons of
zero size: to regularize its behavior one needs to introduce
an ultraviolet cutoff. We have studied in detail the statistical mechanics
with varying cutoff parameter: the numerical simulations are
fully compatible with the expectations.
\newline
\newline
At $N_c=3,4$ the system does thermodynamically exist with no cutoff,
however, it exhibits enormous fluctuations which undermine
accurate measurements of physical quantities. Nevertheless,
we have managed to study the instanton size distribution
introducing two methods of its determination,
`geometric' and `lattice'. Using the `lattice' method
with a very fine grid we observe many unphysical small size instantons
for $N_c=3$, which stem from the instability of the product Ansatz.
Though this effect is believed to disappear when the system approaches the
true statistical equilibrium it is very difficult to get rid of it
in practice because of gigantic fluctuations in the system.
The effect is even more pronounced for the flat measure but tends
to vanish when $N_c$ is increased.
\newline
\newline
Though the zindon parametrization of instantons and of their interactions
allows for complete `melting' of instantons and is quite opposite in spirit
to dilute gas Ans\"atze, we observe that zindons, nevertheless, tend to
form `color-neutral' clusters
which can be identified with well-isolated instantons. This
effect is due to a combination of two different factors both
supporting clustering. One factor is the interactions: same-color zindons
are strongly repulsive while different-color zindons are attractive.
The second factor is pure geometry: even with a flat measure the
probability to combine $N_c$ zindons into a neutral cluster smaller than the
average separation is quite sizeable. Both
these factors are expected to be even stronger in four dimensions
appropriate for the Yang-Mills instantons.
\vskip .5true cm
\noindent {\bf Acknowledgments}
\newline
Stimulating discussions with Dima Khveschenko, Alan Luther and especially
with Victor Petrov are gratefully acknowledged.
\vskip .5true cm

\newpage

%
% Table 1  density in the Coulomb gas, dependent on the cutoff eps2
%
\begin{table}
\begin{tabular}{|l|r|c|c|r|c|}
\hline &&&&&\\
${\epsilon'}'^2$ & density $n_+$
& average distance $1/\sqrt{n_+}$ &
$L_{\rm max}$ & $N_{\rm max}$ & defect $\alpha$\\
&&&&&\\
\hline
\hline
&&&&&\\
   0.1 &    5.095  $\pm$  0.021 &  0.443 &   1.401  &   10 & 0.839 \\
  0.03 &   12.707  $\pm$  0.107 &  0.281 &   1.620  &   33 & 0.777 \\
  0.01 &   23.501  $\pm$  0.211 &  0.206 &   2.063  &  100 & 0.672 \\
 0.003 &   42.723  $\pm$  0.673 &  0.153 &   2.793  &  333 & 0.571 \\
 0.001 &   64.117  $\pm$  1.193 &  0.125 &   3.949  & 1000 & 0.481 \\
0.0003 &  135.207  $\pm$  4.299 &  0.086 &   4.965  & 3333 & 0.518 \\
&&&&&\\
\hline
\end{tabular}
\caption{Dependence of the zindon density in the Coulomb gas
on the cutoff parameter. The error is given by the standard deviation
of the  $n_+$'s obtained by separate fits to five measured curves.
The value of $n_+$ itself is the value one obtains if one performs
a combined fit with one $n_+$ to all the measured curves. The real
value of $n_+$, however, is determined by the error in the defect
$\alpha$, to gain experience of its influence one may compare to
Tab.~\ref{runcou}.}
\label{tab1}
\end{table}

%
% Tabel 2 running coupling data
%
\begin{table}
\begin{tabular}{|c||r|r|r|}
\hline
&&& \\
$\beta_1$ &${\epsilon'}^2 = 0.1$&${\epsilon'}^2 = 0.01$&${\epsilon'}^2 = 0.001$\\
          &$\alpha = 0.7539$ & $\alpha = .5785$ &  $\alpha =0.45992$ \\
          &$(A=  -0.036 $) & $(A= -0.035)$ & $A=  -0.013$ \\
&&& \\
\hline
\hline
&&& \\
0.0 &6.946 & 26.314 &  74.948 \\
0.1 &7.062 & 27.614 &  75.371 \\
0.2 &6.945 & 27.650 &  75.250 \\
0.3 &7.060 & 28.229 &  74.075 \\
0.4 &7.080 & 27.711 &  81.660 \\
0.5 &7.164 & 27.707 &  77.512 \\
0.6 &7.006 & 29.171 &  80.946 \\
0.7 &7.149 & 28.706 &  82.323 \\
0.8 &7.306 & 30.013 &  91.595 \\
0.9 &7.395 & 31.705 &  93.606 \\
1.0 &7.624 & 34.757 &  119.714 \\
&&& \\
\hline
\end{tabular}
\caption{Dependence of the instanton density $n_+$ on the coupling
parameter $\beta_1$. In the table the density $n_+$ is given
for three different values of ${\epsilon'}^2$ ($L=1$). One should
note that the defect $\alpha$ has
quite a large error, therefore the resulting values for the
density at $\beta_1=0$ are different from the one in Tab.~\ref{tab1}
due to this statistical error. $A$ is the intercept from the combined fit
of the free energy to a form $F = A + N[\ln(N/L^2)-1]+N\ln n+\alpha\ln N!$
and it is a small quantity, which can be neglected in the further discussion.}
\label{runcou}
\end{table}

%
% Table 3 Debye fit to the running coupling data
%

\begin{table}
\begin{tabular}{|l||r|r|}
\hline
&&\\
${\epsilon'}^2$ & $a$ & $b$\\
&&\\
\hline
\hline
&&\\
0.1    &   1.94273  &  0.119835 \\
0.01   &   3.29677  &  0.331814  \\
0.001  &   4.29687  &  0.607095  \\
&&\\
\hline
\end{tabular}
\caption{Fit of the equilibrium density to the form given by the
Debye-H\"uckel approximation. The table shows the results for the linear fit
to the form $\ln n_+ = a + b \ln f(\beta_1)$, where
$\ln f(\beta_1) = \frac{1}{2}
[(1+\beta_1)\ln(1+\beta_1) + (1-\beta_1)\ln(1-\beta_1)]$. The increase
of the intercept when ${\epsilon'}^2$ becomes smaller is due to the fact
the the density for $N_c = 2$ is divergent. All values are for $\beta=1$.}
\label{debfit}
\end{table}
%%
%% Table 4 Density for the running coupling case
%%
%
%\begin{table}
%\begin{tabular}{|l||r|r|}
%\hline
%&&\\
%${\epsilon'}^2$ & $n_+$ & $\beta_1$\\
%&&\\
%\hline
%\hline
%&&\\
%0.1    &   7.303     &   0.478    \\
%0.01   &    30.084    &  0.428     \\
%0.001  &    83.989   &   0.323    \\
%&&\\
%\hline
%\end{tabular}
%\caption{Resulting density for the case of a running
%$\beta_1 = \frac{1}{2}\left[\ln\frac{N_+}{L^2}+C\right]$. The result depends
%on the constant $C$, which cannot be determined completely by the
%theory. A workable procedure is to use $C = 1-\ln(n_+_0)$, where
%$n_+_0 = n_+(\beta_1=1)$ for a fixed $\beta_1$}
%\label{runbeta}
%\end{table}
%
%
%\begin{table}
%\begin{tabular}{|c||c|c|c|c|}
%\hline
%&&&&\\
% $N_c$ & $n_+_0$ &
%         $n_+_{\rm defect}$ &
%         defect $\alpha$     &
%         $n_+_{\rm Debye}(\beta=1,\beta_1=0.3)$  \\
%&&&&\\
%\hline
%\hline
%&&&&\\
%3  &    7.3124 &  985.8043 &  1.86611 &   2.934  \\
%&&&&\\
%4  &    0.0427 &   38.3563 &  2.58784 &  2.122   \\
%&&&&\\
%\hline
%\end{tabular}
%\caption{Average instanton densities obtained by various methods:
%$n_+_0$ gives
%the density assuming that the system is stable.
%$n_+_{\rm defect}$ is the
%density, when we fit in addition a defect parameter $\alpha$.
%$n_+_{\rm Debye}$ gives the density predicted
%by the Debye-H\"uckel approximation.
%As the density is averaged over $\beta_1$ ranging from $\beta_1=0.0$ to
%$\beta_1=0.6$, we took the Debye-H\"uckel density at an average
%value of $\beta_1=0.3$.
%Note that $n_+= N_+/V$.}
%\label{dens}
%\end{table}

% figure 1

%

% figure 2

%
% figure 3  snapshots

% figure 4 divergence/nondivergence

% figure 5

%
% figure 6

%
%

%
% Figure 7
%

%

% figure 8: densities

% figure 9: Debye densities

% figure 10: Size distribution with the flat measure

% figure 11: Size distribution with the full measure

% figure 12: Illustration

\begin{figure}
\centerline{\psfig{figure=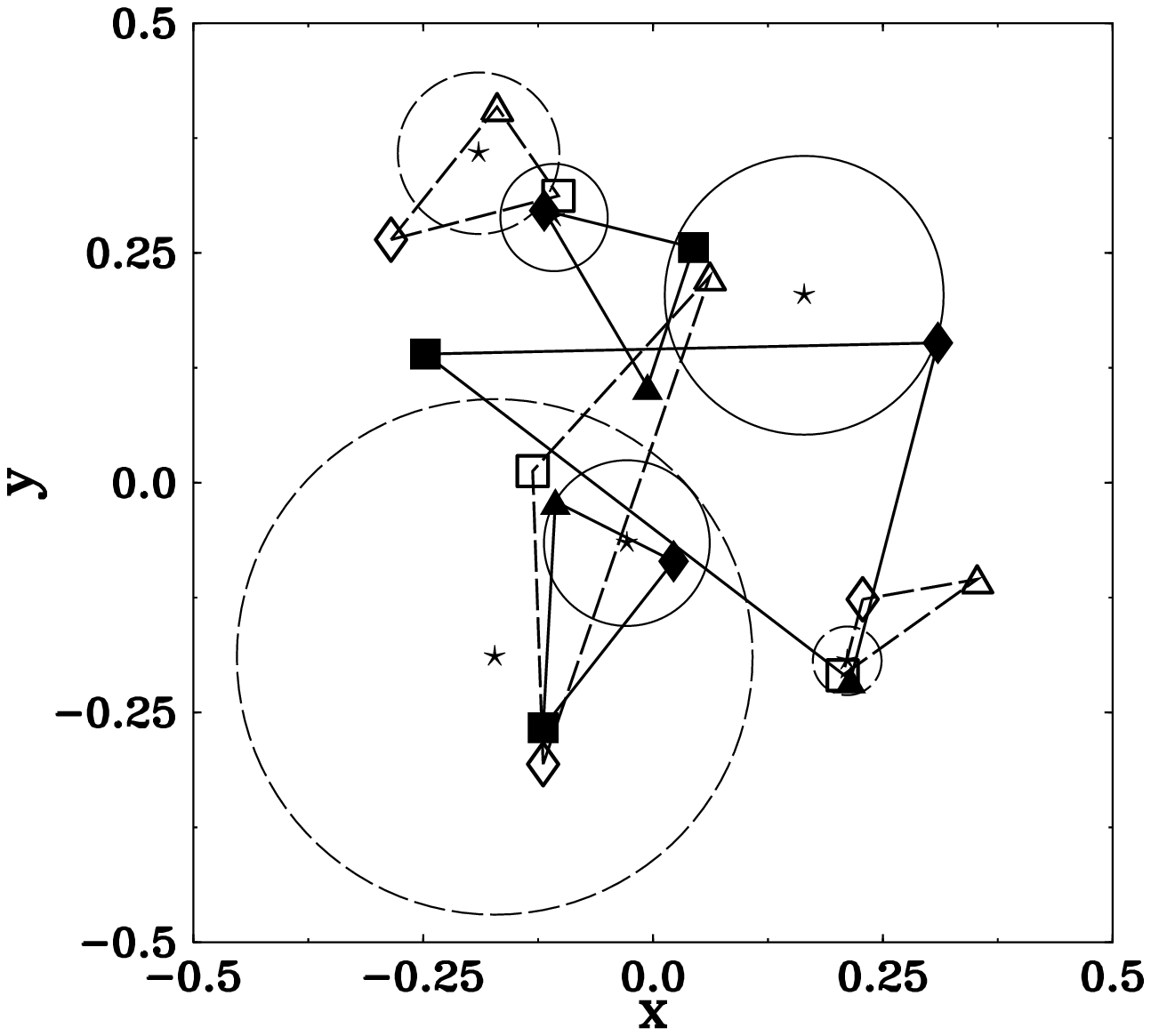,width=8cm}
            \psfig{figure=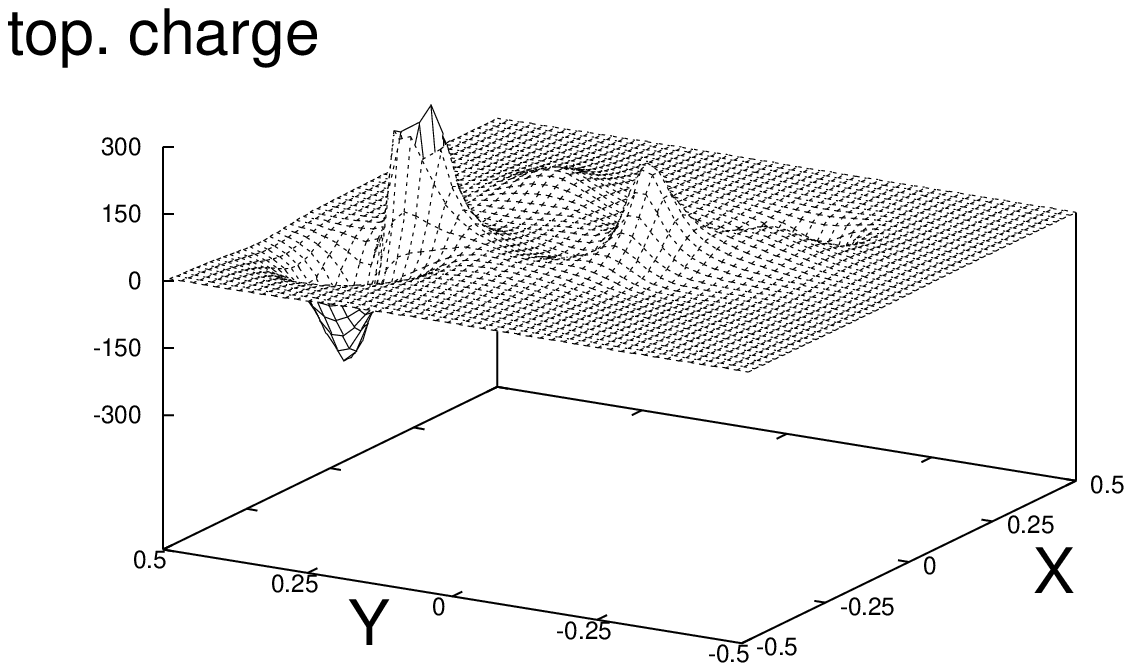,width=8cm}}
 \caption{
Topology of
a sample with 3 instantons and 3 anti-instantons ($N_c=3$) seen by the
`lattice' method. The sample has been generated with the full
partition function at the plateau for $\beta_1 = 0.5$. On the left we show
the the zindon content of the configuration. The triangles connect zindons of three different
colors forming instantons (solid lines) and anti-instantons (dashed lines),
as found by the `geometric' method. Filled symbols correspond to zindons,
open symbols to anti-zindons. The diamonds, squares and triangles denote
the different colors. The stars show the centers and the circles
show instanton sizes, as found by the `lattice' method using a
$100\times 100$ grid. The method has been applied to the
topological charge density displayed as a 3-dimensional plot on the right.}
\label{illustration}
\end{figure}


\newpage
\begin{thebibliography}{99}

\bibitem{Ne98}
J.~W.~Negele, \NPS{73}{92}{1999}. 

\bibitem{SS98}
T.~Schaefer and E.~Shuryak, \RMP{70}{323}{1998}.

\bibitem{DP86}
D.~I.~Diakonov, V.~Yu.~Petrov, \NPB{272}{457}{1986}, \\
D.~Diakonov, in: {\it Selected Topics in Non-perturbative QCD},
Proc. Enrico Fermi School, Course CXXX,
A.~Di Giacomo and D.~Diakonov, eds., Bologna, (1996) p.397,
{\tt hep-ph/9602375}.

\bibitem{GP}
V.~L.~Golo and A.~M.~Perelomov, \PLB{79}{112}{1978}.

\bibitem{ALV78}
A.~D'Adda, M.~Luescher and P.~Di Vecchia, \NPB{146}{63}{1978}.

\bibitem{FFS79a}
V.~Fateev, I.~Frolov and A.~Schwartz, \NPB{154}{1}{1979}.

\bibitem{FFS79b}
V.~Fateev, I.~Frolov and A.~Schwartz, \SNP{30(4)}{590}{1979}.

\bibitem{BL79}
B.~Berg and M.~L\"uscher, \CMP{69}{57}{1979}.

\bibitem{Wi79}
E.~Witten, \NPB{149}{285}{1979}.

\bibitem{FR86}
L.~D.~Faddeev, N.~Yu.~Reshetikhin, \AOP{167}{227}{1986}.

\bibitem{Vi96}
M.~Campostrini, A.~Pelissetto, P.~Rossi, E.~Vicari \PRD{54}{1782}{1996}.

\bibitem{Vi92}
M.~Campostrini, P.~Rossi, E.~Vicari, \PRD{46}{4643}{1992}.

\bibitem{Vi92a}
A.~Di Giacomo, F.~Farchioni, A.~Papa, E.~Vicari, \PRD{46}{4630}{1992}.

\bibitem{MS94}
C.~Michael, P.~S.~Spencer, \PRD{50}{7570}{1994}.

\bibitem{BuLi}
A.~P.~Bukhvostov and L.~N.~Lipatov, \NPB{180}{116}{1981};
Pisma Zh.~Eksp.~Teor.~Fiz. {\bf 31}, 138 (1980).

\bibitem{BP75}
A.~A.~Belavin, A.~M.~Polyakov,
JETP Lett. {\bf 22}, 245 (1975);
Pisma Zh.~Eksp.~Teor.~Fiz. {\bf 22}, 503 (1975).

\bibitem{DP}
D.~Diakonov and V.~Petrov, {\em Towards a New Formulation of the Yang--Mills
Theory}, report PNPI-90-1581 (1990), unpublished.

\bibitem{Jev}
A.~Jevicki, \NPB{127}{125}{1977}.

\bibitem{LLbook} L.~D.~Landau and E.~M.~Lifshitz, Statistical Physics,
chapter VII,p.~74, Pergamon Press, London-Paris, 1958.

\bibitem{Jev2}
A.~Jevicki,  \PRD{20}{3331}{1979}.

\bibitem{Polbook}
A.~Polyakov, {\em Gauge Fields and Strings}, Harwood Academic Publishers,
1987.

\bibitem{ACEP99}
B.~Alles, L.~Cosmai, M.~D'Elia, A.~Papa,
{\em Topology in the $CP^{N-1}$ models: a critical comparision of
different cooling techniques},
BARI-TH-355-99, Sep 1999. 3pp., e-Print Archive: hep-lat/9909034.


%\bibitem{BL81}
%B.~Berg, M.~L\"uscher, \NPB{190}{412}{1981}.
%
%\bibitem{Lu82}
%M.~L\"uscher, \NPB{200}{61}{1982}.

\end{thebibliography}
\end{document}